\documentclass[a4paper,11pt]{article}
\pdfoutput=1

\usepackage{jcappub}

\usepackage[T1]{fontenc}

\usepackage{graphicx}
\usepackage{enumitem}
\usepackage{latexsym}
\usepackage{amsfonts}
\usepackage{amssymb}
\usepackage[dvipsnames]{xcolor}
\usepackage{amsmath}
\usepackage{slashed}
\usepackage{dcolumn}
\usepackage{verbatim}
\usepackage{float}
\usepackage{multirow}
\usepackage{xspace}
\usepackage[normalem]{ulem}
\usepackage[caption=false]{subfig}
\usepackage{bm}
\usepackage{bbm}

\newcommand{\Mpl}{M_{\rm pl}}
\newcommand{\vev}[1]{\left\langle#1\right\rangle}

\allowdisplaybreaks

\begin{document}

\title{
Unified dark sector approaches to cosmological tensions
}

\author{Sergio Sevillano Mu\~{n}oz,} 
\emailAdd{sergiosm@sas.upenn.edu}
\affiliation{Center for Particle Cosmology, Department of Physics and Astronomy, University of Pennsylvania, Philadelphia, Pennsylvania 19104, USA}
\author{Mark Trodden} \emailAdd{trodden@upenn.edu}

\abstract{
We present an analytical study of minimal mass-varying dark matter models, in which a single scalar field controls the dark matter mass, focusing on their utility in simultaneously addressing multiple cosmological tensions. We show that such minimal models can naturally become relevant at two critical epochs in the history of the Universe; matter-radiation equality and the present day, generating an early dark energy-like energy injection, as motivated by the Hubble tension, and an apparent phantom crossing, as suggested by baryon acoustic oscillation data from the DESI project. 
Crucially, the characteristic timing of these effects is tied to the cosmological background evolution itself, rather than specific energy scales.
In addition, the intermediate evolution reduces the clustering matter density and can therefore lower $S_8$, in the direction preferred by data, while avoiding new long-range interactions. We derive analytic arguments to understand the conditions for these effects to arise in the same model and assess whether they can occur without introducing additional fine-tuning. This provides a simple framework for identifying which potentials and couplings to dark matter can simultaneously affect early- and late-time cosmological tensions.
}

\maketitle

\section{Introduction}
It is a remarkable success of modern cosmology that the $\Lambda$CDM model remains a robust fit to our increasingly detailed datasets. As an important consequence, a fundamental understanding of the microscopic natures of dark matter and dark energy, their possible interactions, and the relation of this {\it dark sector} to the visible one are among the largest outstanding questions in our current understanding of the Universe. However, recent cosmological observations may be opening a new window onto the dynamics of these mysterious entities.

One piece of evidence arises from tensions between cosmological parameters obtained from different, independent measurements. The most prominent example is the Hubble tension~\cite{Verde:2019ivm,DiValentino:2020zio,Sekiguchi:2020teg,DiValentino:2021izs,Schoneberg:2021qvd,Kamionkowski:2022pkx,CosmoVerseNetwork:2025alb,Cai:2026swf}, emerging originally from the discrepancy between the value of the expansion rate inferred from Planck CMB observations assuming $\Lambda$CDM ($H_0 = 67.4 \pm 0.5\,{\rm km\,s^{-1}\,Mpc^{-1}}$)~\cite{Planck:2018vyg,ACT:2023dou,ACT:2023kun}, and the local distance-ladder measurement ($H_0 = 73.04 \pm 1.04\,{\rm km\,s^{-1}\,Mpc^{-1}}$) from the SH0ES collaboration~\cite{Riess:2021jrx}, but now reinforced by a growing set of independent observations~\cite{H0LiCOW:2019pvv,Freedman:2020dne,Birrer:2020tax}. This has motivated models that modify the expansion history mainly before recombination,\footnote{While it is theoretically possible to also address the Hubble tension with late time modifications, this approach is not very efficient at fully resolving the tension~\cite{Cai:2021weh,Bansal:2026axl}.} for example through an energy injection at matter-radiation equality that reduces the sound horizon and thereby increases the inferred value of $H_0$~\cite{Pettorino:2013ia,Kamionkowski:2014zda,Poulin:2018cxd,Sakstein:2019fmf,Alexander:2019rsc,Smith:2019ihp,Lin:2019qug,Gogoi:2020qif,Karwal:2021vpk,Niedermann:2021ijp,Copeland:2023zqz}. More recently, baryon acoustic oscillation (BAO) measurements from the DESI project~\cite{DESI:2024mwx,DESI:2025zgx}, when combined with CMB and supernova data, have also been interpreted as providing evidence for evolving dark energy in some parameterizations. In particular, $w_0w_a$-type fits seem to favor an apparent crossing of the phantom divide at small redshifts~\cite{Zheng:2024qzi,Cortes:2024lgw,Orchard:2024bve,Shlivko:2024llw,Notari:2024rti,Fikri:2024klc,Giare:2024oil,DESI:2025fii,Lee:2025pzo,Pan:2025qwy,Scherer:2025esj,Toomey:2025xyo,Nesseris:2025lke,Shlivko:2026jxa}, leading to approaches using standard quintessence fields~\cite{Copeland:1997et,Copeland:2006wr,Burgess:2021obw,Ramadan:2024kmn,Bhattacharya:2024hep,Roy:2025cxk,Shlivko:2025fgv,Caldwell:2025inn,deSouza:2025rhv,Shajib:2025tpd,Jiang:2026cqh}. However, it is well-known that phantom dark energy models alone face a formidable set of theoretical obstacles~\cite{Carroll:2003st,Cline:2003gs,Dubovsky:2005xd,Nicolis:2009qm, Creminelli:2010ba,Creminelli:2012my}, and hence it is important to explore alternative theoretical solutions.

Among the possible extensions of $\Lambda$CDM proposed to address cosmological tensions, an interacting dark sector provides a particularly well-motivated framework, partly because it can be easily connected to the relevant cosmological epochs~\cite{Amendola:1999er, Brax:2004qh, Amendola:2006dg}. While these models have been well explored in the early-Universe regime through early-dark-energy (EDE) scenarios~\cite{Doran:2006kp,Karwal:2021vpk,Alexander:2022own,Lin:2022phm,Bernui:2023byc,Pan:2023mie,Liu:2023rvo,Pitrou:2023swx,Zhai:2023yny,Smith:2024ibv,Kamionkowski:2024axz,Smith:2024ayu,Smith:2025grk,Smith:2025uaq}, they are especially effective at fitting the late-time expansion of the Universe~\cite{Huey:2004qv,DiValentino:2017iww,DiValentino:2019jae,vandeBruck:2022xbk,Poulot:2024sex,Giare:2024smz,Ye:2024ywg,Wolf:2024eph,Teixeira:2024qmw,Khoury:2025txd,Wolf:2025jlc,Poulin:2025nfb,Yashiki:2025loj,Pourtsidou:2025sdd,Linder:2025zxb,Zhai:2025hfi,Yao:2025wlx,Smith:2025icl,SanchezLopez:2025uzw,Liu:2025bss,Toomey:2025yuy,Figueruelo:2026eis,Jensko:2026taf,Wang:2026wrk,Antusch:2026ldp,Calderon:2026hbr,Teixeira:2026yjd,Naidoo:2026umv,Delaunay:2026jto,Garcia-Garcia:2026nzy,Lee:2026yzs,Li:2026xaz,Khoury:2026svx}. The crucial point here is that the evolution of the dark matter density in these models differs from the simple evolution in uncoupled models. As a result, if observations are interpreted assuming standard cold dark matter (CDM) behavior, the inferred behavior of the dark-energy component can differ from the physical one. In particular, if the effective dark matter mass increases at low redshift, the mismatch between the true and inferred dark matter evolutions can lead to an apparent phantom crossing for dark energy, even when no fundamental field violates the null energy condition~\cite{Carroll:2004hc,Das:2005yj,Brookfield:2005bz,Agrawal:2019dlm,Bedroya:2025fwh}. This effect is often referred to as a \emph{phantom mirage}, since the inferred dark energy appears phantom-like while the underlying theory remains non-phantom.

While a large literature has studied these models as independent explanations of either early- or late-time deviations from $\Lambda$CDM, there is growing interest in exploring whether both effects can arise simultaneously~\cite{Freese:2021rjq,Adil:2022hkj,Ramadan:2023ivw, Sohail:2024oki,Wang:2024dka,Baidya:2026tyh}. Recently, Ref.~\cite{Giare:2026tyk} showed that this can be achieved if the scalar-field potential contains at least three regions with different slopes. In this work, we focus instead on a minimal setup with a single bare potential and a coupling to dark matter~\cite{Casas:1991ky,Garcia-Bellido:1992xlz,Anderson:1997un,Amendola:2000uh,Farrar:2003uw,Fardon:2003eh,Bean:2008ac,Brax:2011ja,Wang:2016lxa,vandeBruck:2019vzd,Trodden:2022zye,Mandal:2022yym,Das:2023enn,SevillanoMunoz:2024ayh,Khoury:2025txd}.  We develop analytic arguments to identify the general conditions under which such interactions can address both early- and late-time cosmological tensions while satisfying theoretical stability requirements from effective field theory considerations. This allows us to formulate model-building criteria for viable scenarios and to assess the degree of fine-tuning required to achieve them.

A notable feature of this construction is that the two relevant epochs do not need to be introduced as independent scales in the scalar sector. At early times, the coupling-induced force is weighted by the matter abundance and therefore becomes important naturally near matter-radiation equality. At late times, for exponential or slowly varying-slope potentials, the scalar leaves slow roll when its own energy density becomes cosmologically relevant. The timing of both effects is therefore linked directly to the background cosmological evolution, while their sizes remain controlled by the dark matter mass function and the scalar potential.

In Section~\ref{sec:Dark interactions} we show how interactions in the dark sector can give rise to an effective dark energy description capable of producing both early-time energy injections and late-time phantom crossing. In Section~\ref{sec:MReq}, we focus on the early-Universe regime and derive an analytic estimate for the resulting early dark energy peak amplitude. In Section~\ref{sec:DEMeq}, we apply the same framework to the late Universe, where dark energy becomes cosmologically relevant, and assess the fine-tuning required to fit DESI BAO-like data. We conclude in Section~\ref{sec:Conclusion}.

\section{Dark energy from dark interactions}\label{sec:Dark interactions}
While many interactions between dark matter and dark energy are possible, we focus here on the minimal case in which a quintessence field couples directly to the dark matter mass. The dark-sector equations of motion take the form
\begin{align}\label{eq:phi eom}
\ddot{\phi}+3H\dot{\phi}=-V_{,\phi}(\phi)-\frac{\mathrm{d}\ln m(\phi)}{\mathrm{d}\phi}\rho_{\rm dm}(\phi),\\
{\rho}_{\rm dm}(\phi)=\rho_{{\rm dm},i}\frac{m(\phi)}{m(\phi_i)}\left(\frac{1+z}{1+z_i}\right)^3,
\end{align}
where $m(\phi)$ is the field-dependent mass of the dark matter component, and $\rho_{\rm dm}(\phi)$ is its corresponding energy density. Notice that in writing these equations, we remain agnostic about the microscopic nature of dark matter. This is because independently of whether the underlying dark matter degrees of freedom are fermionic or bosonic, their cosmological evolution can be described by the same effective fluid equations.

In the absence of interactions, one would usually identify the dark energy density with that of the quintessence field, $\rho_\phi=\dot{\phi}^2/2+V(\phi)$. However, in the presence of interactions this identification becomes more subtle. This is because the dark matter energy density no longer redshifts purely as $a^{-3}$, since the evolution of $m(\phi)$ induces an additional contribution to its time dependence. This simply corresponds to a dark-sector energy exchange, but it can have important consequences for how we interpret observations if the expansion history is inferred under the incorrect assumption that dark matter is an uncoupled pressureless fluid. In that case, the non-standard evolution of $\rho_{\rm dm}(\phi)$ is absorbed into an effective dark energy component, which can be defined as 
\begin{equation}\label{eq:rho de}
\rho_{\rm de}=\rho_\phi+\rho_{\rm dm}-\rho_{\rm dm,0}a^{-3},
\end{equation}
where $\rho_{\rm dm,0}$ is the dark matter density measured today and, since we are assuming a flat FRW metric, we set $a=a_0=1$ today. From this, one can already make explicit the connection with the phantom-mirage mechanism introduced above. The effect appears when one computes the effective equation of state associated with $\rho_{\rm de}$, which is given by
\begin{equation}\label{eq: weff}
w_{\rm de}=\frac{\frac{1}2\dot{\phi}^2-V(\phi)}{\frac{1}2\dot{\phi}^2+V(\phi)+\left[\frac{m(\phi)}{m(\phi_0)}-1\right]{\rho_{\rm dm,0}a^{-3}}},
\end{equation}
where $m(\phi_0)$ is the present-day dark matter mass. From this expression, we see that if $m(\phi(z>0))<m(\phi_0)$, the effective dark energy equation of state can cross the phantom divide without requiring any fundamental phantom-like field.\footnote{It is also possible to  infer an apparent crossing of the phantom divide with a decreasing dark matter mass if $\rho_{\rm de}$ is measured at a higher redshift $z_{\rm obs}$, such that $\rho_{\rm de}(z_{\rm obs})=\rho_{\phi}(z_{\rm obs})$~\cite{Agrawal:2019dlm,Bedroya:2025fwh}.}

Equations (\ref{eq:rho de}) and (\ref{eq: weff}) already provide useful intuition for how this minimal model can address both early- and late-time cosmological tensions. The interaction term, $m(\phi)$, affects cosmological evolution in two distinct ways. First, the field-dependent dark matter mass contributes to the equation of motion for $\phi$, modifying the scalar field evolution when the dark matter abundance becomes important, around matter-radiation equality. Second, the same mass dependence changes the evolution of dark matter when the $\phi$ field evolves. If this effect is interpreted assuming standard cold dark matter, the resulting mismatch is absorbed into an effective dark energy component and can contribute both to an early-time energy injection and to an apparent phantom crossing at late times.

We will begin by studying how these fields can produce an injection of energy around matter-radiation equality, which sets the initial conditions for the subsequent phantom behaviour at late times.

\section{Early time: Matter-Radiation equality}\label{sec:MReq}
The dominant contribution to the evolution of $\phi$ at this epoch comes from its interaction with dark matter. This is because the bare potential $V(\phi)$ is chosen to drive the accelerated expansion of the Universe only at late times, and is therefore subdominant around matter-radiation equality. We can then approximate the equation of motion for $\phi$ as
\begin{equation}\label{eq:phirhodm}
    \ddot{\phi}+3H\dot{\phi}=-\frac{\mathrm{d}\ln m(\phi)}{\mathrm{d}\phi}\rho_{\rm dm}(\phi),
\end{equation}
where
\begin{equation}
    H^2=\frac{1}{3\Mpl^2}\left(\rho_{\rm r}+\rho_{\rm b}+\rho_{\rm dm}(\phi)+\frac{1}{2}\dot{\phi}^2\right),
\end{equation}
with $\rho_{\rm r}$ and $\rho_{\rm b}$ being the radiation and baryon energy densities, respectively. Although the explicit $\phi$-dependence of $\rho_{\rm dm}$ may appear to complicate the solution of Eq.~\eqref{eq:phirhodm}, the system becomes simpler when written in terms of derivatives with respect to the number of e-folds, $N=\log a$. This gives
\begin{equation}
    {\phi''}+\left(3+\frac{H'}{H}\right){\phi}'=-3\Mpl^2\frac{\mathrm{d}\ln m(\phi)}{\mathrm{d}\phi}\Omega_{\rm dm}(\phi),
\end{equation}
where $\Omega_{\rm dm}=\rho_{\rm dm}/3\Mpl^2H^2$ is the energy fraction of the dark matter fluid. However, as discussed in Section~\ref{sec:Dark interactions}, the most convenient quantity to use is not the physical dark matter fraction $\Omega_{\rm dm}(\phi)$, but the inferred matter fraction
\begin{equation}
\Omega_{\rm m}=\Omega_{\rm b}+\Omega_{\rm CDM},
\end{equation}
where
\begin{equation}
\Omega_{\rm CDM}
\equiv
\Omega_{\rm dm}(\phi)\frac{m(\phi_0)}{m(\phi)}=\frac{\rho_{\rm dm,0}(1+z)^3}{3\Mpl^2H^2}
\end{equation}
is the dark matter fraction that would be inferred under the assumption of standard cold dark matter. With this, the equation of motion takes the form
\begin{equation}\label{eq:phi' OmegaM}
    {\phi''}+\left(3+\frac{H'}{H}\right){\phi}'=-3\Mpl^2\frac{m_{,\phi}(\phi)}{m(\phi_0)}\Omega_{M}(1-r),
\end{equation}
with $r\equiv\Omega_{\rm b}/\Omega_{\rm m}\approx0.15$. This equation makes the physical origin of the effect transparent. For a fixed initial value of $\phi$, the interaction is suppressed when $\Omega_{\rm m}$ is small, but becomes important as the Universe approaches matter-radiation equality. Around this epoch, the coupling to dark matter can drive the field away from slow roll and, depending on whether $m(\phi)$ has a minimum, induce a peak in the effective dark energy density. 

The corresponding inferred dark energy fraction around this period is
\begin{equation}
    \Omega_{\rm de}=\Omega_{\phi'}+ \left(\frac{m(\phi)}{m(\phi_0)}-1\right)(1-r)\Omega_{\rm m},
\end{equation}
where $\Omega_{\phi'}\equiv \phi'^2/(6\Mpl^2)$ is the fractional kinetic energy of the $\phi$ field. From this expression, the evolution of the effective dark-energy fluid can be understood in terms of three distinct regimes:
\begin{itemize}
    \item \textbf{The slow-roll regime:} At early times the field can remain frozen close to its initial value as Hubble friction dominates over the influence of the potential term. To see how long this lasts, we linearize the field equation, writing $\phi=\phi_i+\delta\phi$, giving
    \begin{equation}
    \delta\phi''+\left(3+\frac{H'}{H}\right)\delta\phi'
    =
    -3\Mpl^2
    \frac{m_{,\phi\phi}(\phi_i)}{m(\phi_0)}
    \Omega_{\rm m}(1-r)\delta\phi .
    \end{equation}
    Therefore, perturbations around the initial field value are strongly damped whenever the effective mass induced by the coupling is small compared to the Hubble scale,
    \begin{equation}
    \left|
    3\Mpl^2
    \frac{m_{,\phi\phi}(\phi_i)}{m(\phi_0)}
    \Omega_{\rm m}(1-r)
    \right|
    <1 .
    \end{equation}
    During this regime the field is approximately constant, $\phi\simeq\phi_i$, such that $\Omega_{\phi'}=0$ and the effective dark-energy fraction evolves as
    \begin{equation}\label{eq:DE_peak}
    \Omega_{\rm de}
    \approx
    \left(
    \frac{m(\phi_i)}{m(\phi_0)}-1
    \right)
    (1-r)\Omega_{\rm m} .
    \end{equation}
    We can see that $\Omega_{\rm de}$ naturally grows with time during radiation domination, following the growth of $\Omega_{\rm m}$ towards matter-radiation equality. This produces an EDE-like contribution, although the effective fluid does not initially have $w_{\rm de}\simeq -1$, unlike most conventional EDE models. This is not a problem by itself, as this would still enhance the expansion rate before recombination.
    \item \textbf{The peak:} Once the slow-roll condition breaks down, the field begins to move towards the minimum of the effective potential induced by the dark-matter coupling. If this transition takes place close to matter-radiation equality, where $\Omega_{\rm m}\simeq1/2$, then the breakdown of slow-roll implies
    \begin{equation}
    3\Mpl^2 \frac{m_{,\phi\phi}(\phi_i)}{m(\phi_0)} (1-r) \simeq 2 . 
    \end{equation}
    Substituting this into Eq.~\eqref{eq:DE_peak} gives a peak amplitude of approximately
    \begin{equation} 
    \Omega_{\rm de}^{\rm max} \approx \mathcal{O}\left(\frac{1-r}{2} \left( \frac{m(\phi_i)}{m(\phi_0)}-1 \right)\right),
    \end{equation}
    where $1-r\simeq0.8$, and where the early evolution of $\phi$ should be taken into account to get a more precise estimate. Therefore, the time at which the peak occurs is controlled mainly by the curvature ratio $m_{,\phi\phi}(\phi_i)/m(\phi_0)$, while the height of the peak is controlled mainly by the mass ratio $m(\phi_i)/m(\phi_0)$. This separation is useful, as these two degrees of freedom can be independently adjusted to fit the observed phenomenology. The only important requirement is that a positive early-dark energy amplitude enforces $ m(\phi_i)>m(\phi_0)$, which determines how the effective fluid must behave after the peak.
    
    \item \textbf{After the peak:}  
    Once the field becomes dynamical, its evolution is governed by the shape of $m(\phi)$. From Eq.~\eqref{eq:phi' OmegaM}, once $\Omega_{\rm m}$ becomes order unity, the coupling-induced force remains active unless the field reaches a region where $m(\phi)$ stops evolving, or until another component dominates the expansion, leading to $\Omega_{\rm m}\to0$. Since the EDE-like contribution must disappear before recombination, the simplest viable possibility is for $m(\phi)$ to have a minimum where the field can settle before then. This has the additional benefit of avoiding the generation of new long-range fifth forces between dark matter particles\footnote{Although the strongest fifth-force bounds typically arise from searches for new long-range forces between baryons~\cite{Hees:2018fpg,Lee:2020zjt}, analogous constraints also exist for dark-sector fifth forces~\cite{Kesden:2006vz,Archidiacono:2022iuu}.} while the field rests at the minimum of $m(\phi)$.

    To estimate how the energy density of the effective dark energy contribution redshifts as the field approaches this minimum, we compute its averaged equation of state after the peak. Ignoring the bare potential at this stage, this is given by
    \begin{equation}\label{eq:wde after peak}
        w_{\rm de}
        =
        \frac{\frac12\dot\phi^2}
        {\frac12\dot\phi^2
        +
        \left(\frac{m(\phi)}{m(\phi_0)}-1\right)\rho_{\rm CDM}}.
    \end{equation}
    Assuming the mass term, $m(\phi)$, to have the generic form
    \begin{equation}
        m(\phi)
        =
        m_0
        \left(
        1+\frac{\phi^{2n}}{M^{2n}}
        \right),
    \end{equation}
     we can see that the field will oscillate around the minimum until $m(\phi)\to m_0$. Since the oscillation period is much shorter than a Hubble time, we can treat $\rho_{\rm CDM}$ as approximately constant over one oscillation and apply the virial theorem to the field-dependent part of the effective potential. This gives
\begin{equation}
    \left\langle\frac12
    \dot\phi^2
    \right\rangle
    =
    n
    \left\langle
    \frac{\phi^{2n}}{M^{2n}+\phi_0^{2n}}
    \right\rangle
    \rho_{\rm CDM}.
\end{equation}
    Substituting this relation into Eq.~\eqref{eq:wde after peak}, and after some algebra, we obtain
    \begin{equation}\label{eq:wde virialised}
        \left\langle w_{\rm de}\right\rangle
        =
        \frac{nR}{(n+1)R-1},
    \end{equation}
    where we have defined
    \begin{equation}
        R
        =
        \left\langle
        \frac{\phi^{2n}}{\phi_0^{2n}}
        \right\rangle.
    \end{equation}
    Notice that Eq.~\eqref{eq:wde virialised} differs from the usual result for a scalar oscillating in a power-law potential~\cite{Turner:1983he}, $\left\langle w_\phi\right\rangle=({n-1})/({n+1})$, because the interaction term is proportional to the pressureless dark-matter density, and the inferred dark-energy density contains the additional subtraction of the inferred CDM component. The later evolution of $\rho_{\rm de}$ after the peak depends on the value of $\phi_0$, as for this choice of $m(\phi)$ its average is given by
    \begin{equation}
        \vev{\rho_{\rm de}}=\vev{\frac{1}{2}\dot{\phi}^2}+\left(\frac{\vev{\phi^{2n}}-\phi_0^{2n}}{M^{2n}+\phi_0^{2n}}\right)\rho_{\rm CDM}=[(1+n)R-1]\frac{\phi_0^{2n}\rho_{\rm CDM}}{M^{2n}+\phi_0^{2n}}.
    \end{equation}
    If $\phi_0\to0$, the effective fluid redshifts with $w_{\rm de}\to n/(n+1)$, and therefore decays faster than radiation for any positive integer $n$. However, if instead $\phi_0>0$, as can occur when a bare potential $V(\phi)$ controls the late-time evolution, the dynamics are qualitatively different. For a short period, while $R\gg1$, the effective fluid redshifts with $w_{\rm de}\to n/(n+1)$ as in the $\phi_0\to0$ case. However, as the oscillation amplitude decreases, so does $R$. Once $R=1/(n+1)$, the averaged effective density crosses zero, and for smaller $R$ the inferred $\rho_{\rm de}$ becomes negative and $w_{\rm de}\to0$, so that the effective component behaves as a negative matter-like contribution.
\end{itemize}

\begin{figure}
    \centering
    \includegraphics[width=1\linewidth]{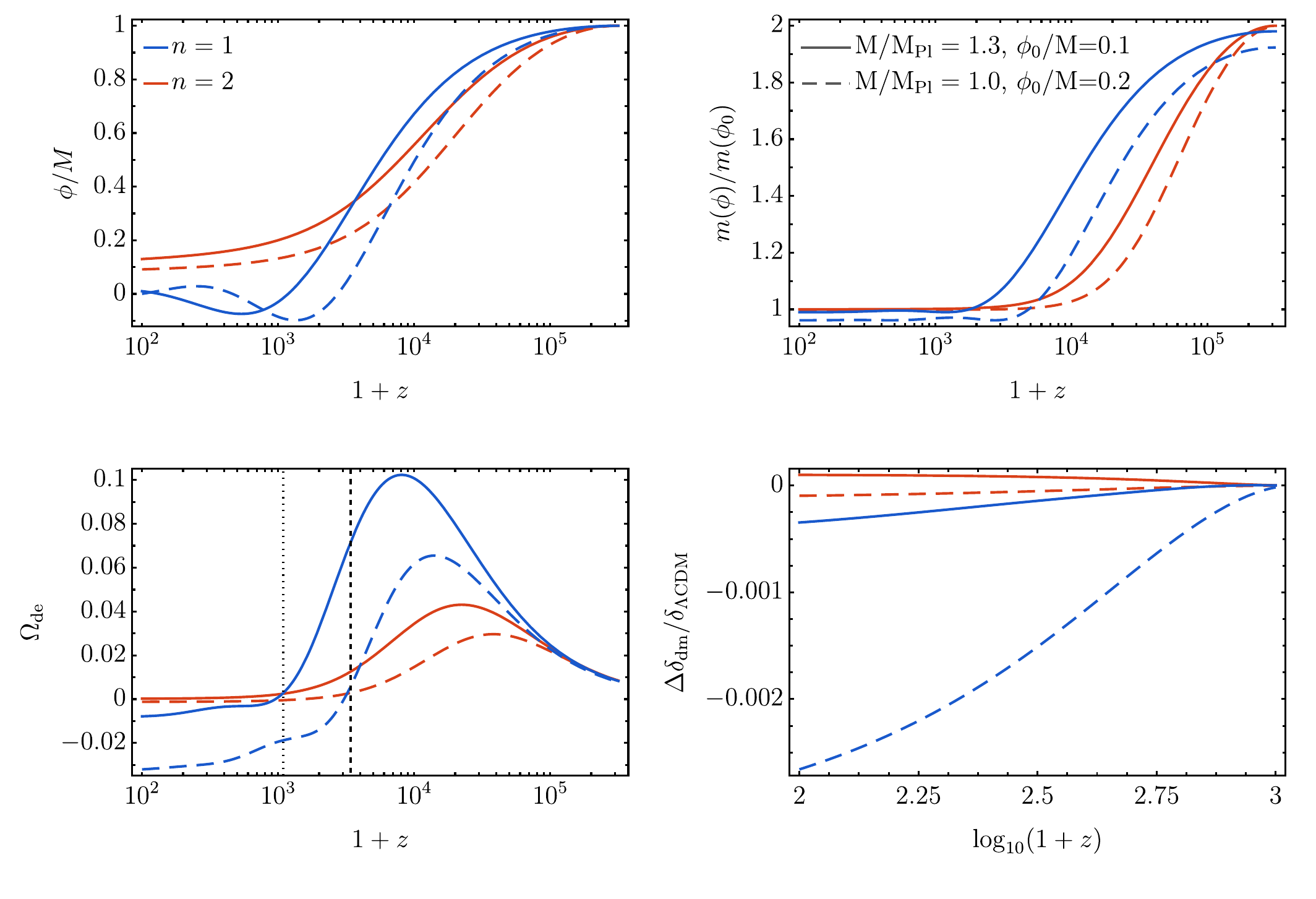}
    \caption{Scalar-field evolution for $n=1$ and $n=2$ couplings between dark matter and a quintessence field $\phi$. For each $n$, we show two choices of the mass scale $M$, together with the corresponding present-day value $\phi_0$. For $M\sim M_{\rm Pl}$, the field naturally begins to evolve around matter-radiation equality, marked by the black dashed line, producing a peak in the effective dark-energy fraction $\Omega_{\rm de}$ that can reach $\mathcal{O}(10\%)$ and dissipate before recombination, marked by the dotted line. The same mass variation suppresses the growth of dark matter perturbations, since $m(\phi)<m(\phi_0)$ during the dark ages, as shown by $\Delta\delta_{\rm dm}=\delta_{\rm dm}-\delta_{\Lambda \rm{CDM}}$.}
    \label{fig:mat_rad_kick}
\end{figure}

Figure~\ref{fig:mat_rad_kick} illustrates this full evolution for $n=1$ and $n=2$ for different parameter choices. For peaks around matter-radiation equality, the peak is more prominent for $n=1$. To see this, note that for $n=1$ the mass is constant, and the post peak evolution is controlled mainly by the standard redshift of the energy. However,for $n>1$, the effective mass of $\phi$ induced by the coupling is field dependent, $m_{,\phi\phi}(\phi)\propto\phi^{2(n-1)}$, so the interaction becomes further suppressed as $\phi$ evolves. We can also see that the peak is larger for larger values of $M$, since increasing $M$ weakens the interaction strength and delays the departure from slow roll. This means that the field then remains displaced for longer, allowing the effective dark energy component to grow relative to radiation before it redshifts away.  The peak height also depends on the mass mismatch between the dark ages and today, with larger $m(\phi_i)/m(\phi_0)$ producing a larger effective energy injection. 

Additionally, there is a close connection with the $S_8$ tension~\cite{Heymans:2020gsg,DES:2021wwk}. During the dark ages, before the bare potential becomes dynamically relevant, we have $m(\phi)<m(\phi_0)$. In this case, the physical dark matter density sourcing structure formation is smaller than the density inferred by assuming standard cold dark matter, such that $\rho_{\rm dm}(\phi)<\rho_{\rm dm,0}a^{-3}$. This reduced clustering density can suppress the growth of dark matter perturbations, moving the model in the direction required to lower $S_8$. As a simple diagnostic, one can compare the growth of $\delta_{\rm dm}$ in this background with its $\Lambda$CDM counterpart by solving
\begin{equation}
\delta_{\rm dm}''+\left(2+\frac{H'}{H}\right)\delta_{\rm dm}'-\frac{3}{2}\Omega_{\rm m}^{\rm eff}(a)\delta_{\rm dm}=0 ,
\end{equation}
where
\begin{equation}
\Omega_{\rm m}^{\rm eff}(a)=\frac{\rho_{\rm b}(a)+\rho_{\rm dm}(\phi)}{3M_{\rm Pl}^2H^2}.
\end{equation}
This calculation isolates the effect of the reduced physical dark matter density on growth. A suppression of $\delta_{\rm dm}$ relative to $\Lambda$CDM would therefore provide evidence that the mass-rescaling mechanism acts in the right direction for the $S_8$ tension. This is supported by calculations in similar models~\cite{Khoury:2025txd,Khoury:2026svx,Delaunay:2026jto}. However, a definitive statement in a specific model requires a full numerical study including scalar perturbations and the modified dark matter continuity and Euler equations. Note that there are no long-range fifth forces during the dark ages, since the field sits at a minimum of $m(\phi)$.

Having a reduced dark matter growth therefore implies a negative energy density during the dark ages. However, this does not imply that any fundamental field has negative energy density, as the total dark sector density remains positive. Rather, it arises from splitting the dark sector into an inferred cold dark matter component with mass $m(\phi_0)$ plus an effective dark energy correction. This negative matter-like contribution persists until the bare potential $V(\phi)$ becomes relevant and drives the field towards $\phi=\phi_0$, producing the late-time \emph{phantom mirage} behavior discussed earlier.

\section{Late time: Matter-Dark energy equality}\label{sec:DEMeq}

During the dark ages, $\phi$ is expected to remain fixed at the minimum of $m(\phi)$ until $V(\phi)$ becomes dynamically relevant and moves the field towards its present-day value, $\phi=\phi_0$~\cite{Thomas:2016iav,Planck:2018vyg}. In that case, the cold dark matter density inferred using the late-time mass $m(\phi_0)$ would not match the physical dark matter density in the past, creating an apparent phantom crossing in the dark energy sector.

For this phantom crossing to occur only at late times, one must explain why the field remains approximately constant until low redshifts, around $z\sim 1$ for agreement with DESI BAO data~\cite{DESI:2025fii}. Depending on the form of $V(\phi)$, there are two qualitatively different ways of delaying the field displacement until low redshift. The first is a \emph{sign-flip mechanism}, in which the bare potential opposes the dark matter induced force and eventually destabilizes the fixed point that trapped the field during matter domination. The second is a \emph{slow-roll exit mechanism}, in which the field remains frozen until the curvature of $V(\phi)$ becomes comparable to the Hubble scale, often referred to as \emph{thawing quintessence}~\cite{Caldwell:2025inn,Liu:2025bss,Caldwell:2005tm}. Both possibilities have advantages and drawbacks. The slow-roll exit mechanism is more generic, since it does not require a special alignment between $V(\phi)$ and $m(\phi)$, but it requires the scalar mass to be tuned so that the field starts rolling only at late times. By contrast, the sign-flip mechanism can work for a wider range of scalar masses, but imposes $V_{,\phi}(\phi)\propto -m_{,\phi}(\phi)$ locally. 

In what follows, we analytically explore both scenarios and determine the conditions that $m(\phi)$ and $V(\phi)$ must satisfy in order to generate a sufficiently large field displacement at low redshift. This will allow us to assess the degree of tuning required in each case to produce an apparent phantom crossing while remaining within the regime of validity of the effective field theory (EFT).

\subsection{Sign-flip mechanism}\label{sec:DEMeq signflip}
The only requirement for the sign-flip mechanism is that, locally, the bare potential has the same shape but opposite sign as the coupling to dark matter, $m(\phi)$. To be as generic as possible, let us consider the evolution of the system under power-law functions with two different scales, $\tilde{M}$ and $M$,
\begin{align}
V(\phi)=&\Lambda_{\rm de}^4\left(1-\frac{\phi^{2n}}{ \tilde{M}^{2n}}\right), & m(\phi)=&m_0\left(1+\frac{\phi^{2n}}{M^{2n}}\right),
\label{eq:potentialforsignflip}
\end{align}
such that the equation of motion takes the form
\begin{equation}\label{eq:signflip}
\ddot{\phi}+3H\dot{\phi}-2n\frac{\phi^{2n-1}}{\tilde{M}^{2n}}\left(\Lambda_{\rm de}^4-\alpha\rho_{\rm m}\right)=0,
\end{equation}
where 
\begin{equation}\label{eq:alpha}
\alpha=\frac{\tilde{M}^{2n}}{M^{2n}}\frac{m_0}{m(\phi_0)}(1-r).
\end{equation}
We see that the system has a stable fixed point at $\phi = 0$ whenever $\rho_{\rm m}>\rho_{{\rm m},c}$, where $\rho_{{\rm m},c}= \Lambda_{\rm de}^4/\alpha$ is the critical density at which the effective potential flips its sign. This fixed point becomes unstable once this inequality is violated, which for $\alpha=1$ occurs precisely at matter--dark energy equality. After this transition, the $\rho_{\rm m}$ contribution redshifts and rapidly becomes subdominant. Therefore, for a given critical redshift $z_c$, the field is released with an initial displacement $\phi_i$ and an initial tachyonic mass $m_\phi(\phi_i)>H_c\equiv H(z_c)$.\footnote{One could have $m_\phi(\phi_i)<H_c$, but then the motion of the field would not be triggered by the sign-flip mechanism but by the exit from slow-roll.} Taking derivatives with respect to e-folds, $N=\log(a)$, and neglecting the matter contribution after the transition, we obtain the following equation of motion
\begin{equation}\label{eq: tachyonic full equation}
\phi''+\left(3+\frac{H'}{H}\right)\phi'-2n\frac{\Lambda_{\rm de}^4}{\tilde{M}^{2n}}\frac{\phi^{2n-1}}{H^2}=0.
\end{equation}
For the late-time Universe, we can ignore the radiation energy density, such that
\begin{equation}\label{eq:H2}
H^2=\frac{1}{3\Mpl^2}\left(\frac{1}{2}\dot{\phi}^2+V(\phi)+\rho_{\rm dm}(\phi)+\rho_{\rm b}\right).
\end{equation}
From this equation, we can already estimate how the field evolves once the sign of the effective potential flips. Although the full evolution is difficult to solve analytically, our goal is simpler, as we want to determine how hard it is to obtain a well-behaved field trajectory, along which $\phi$ evolves sufficiently to produce an appreciable dark matter mass rescaling, but not so rapidly that it leaves the EFT regime or that its speed increases so much that $w_{\rm de}>-1/3$, for which the universe would not be accelerating. To do this, we make two conservative assumptions based on the relative size of the Hubble friction, set by the damping term proportional to $\phi'$, and the potential slope, set by $V_{,\phi}/H^2$. The larger the latter is, the faster the field moves.

The key point is that $H$ decreases monotonically, so $H'<0$. First, the largest friction term is obtained by setting $H'=0$. This overestimates the damping in Eq.~$\eqref{eq: tachyonic full equation}$, and therefore makes the field move more slowly than it would in the full system. Second, the potential term is divided by $H^2$, so the effective slope in the dynamical equation grows as $H$ decreases. By fixing $H=H(z_c)=H_c$, we therefore underestimate the tachyonic force after the transition, since $H_c$ is the largest value of $H$ during the subsequent evolution. We will therefore solve the conservative evolution obtained by setting $H'=0$ and $H=H_c$, given by
\begin{equation}\label{eq: tachyonic approx}
\phi''+3\phi'-2n\frac{\Lambda_{\rm de}^4}{\tilde{M}^{2n}}\frac{\phi^{2n-1}}{H_c^2}=0 \ .
\end{equation}
These approximations do not reproduce the exact evolution of $\phi$, but they provide a conservative lower bound for the field displacement described by Eq.~\eqref{eq: tachyonic full equation}. Therefore, if even this conservative solution requires fine tuning in order to remain within the EFT regime, the full solution will require at least as much tuning, since the true evolution has smaller friction and a stronger effective potential force after the transition.

Although the equation of motion is greatly simplified, there is no simple analytic solution for arbitrary $n$. Therefore, we start with $n=1$, which admits an analytic solution, and then extend the conclusions to higher powers in Appendix~\ref{app: n>1}. For $n=1$, the equation of motion becomes
\begin{equation}\label{eq: tachyonic approx n1}
\phi''+3\phi'-\frac{m_\phi^2}{H_c^2}\phi=0,
\end{equation}
with
\begin{equation}
m_\phi^2=2\frac{\Lambda_{\rm de}^4}{\tilde{M}^{2}}.
\end{equation}
The growing mode for $m_\phi\gg H_c$ can be approximated to
\begin{equation}\label{eq:phi(N) n=1}
\phi(N)=\phi_i\exp\left[\frac{-3+\sqrt{9+4m_\phi^2/H_c^2}}{2}\Delta N\right]\approx\phi_i\exp\left(\frac{m_\phi}{H_c}\Delta N\right),
\end{equation}
where $\Delta N$ is the number of e-folds elapsed. Given a transition at $\rho_{{\rm m},c}=\rho_{\rm m,0}(1+z_c)^3\equiv\Lambda^4_{\rm de}/\alpha$, we can estimate $\Delta N$ by
\begin{equation}
    \Delta N=\log\left({1+z_c}\right)=\frac{1}{3}\log\left(\frac{\Lambda^4_{\rm de}}{\alpha\rho_{\rm m,0}}\right)\approx\frac{1}{3}\log\left(\frac{2.3}{\alpha}\right),
\end{equation}
where in the last step we have substituted for the approximate values of $\Omega_{\rm de,0}\approx0.7$ and $\Omega_{\rm m,0}\approx0.3$. Therefore, the field displacement today is bounded from below by
\begin{equation}
    \phi_0>\phi_i \left(\frac{2.3}{\alpha}\right)^{{m_\phi}/{3H_c}}.
\end{equation}
This shows that the evolution is exponentially sensitive to the initial tachyonic mass, given that $\alpha<1$. Thus, for large $m_\phi/H_c$ a large coincidence is required to achieve $-1<w_{\rm de}<0$, as DESI seems to imply. This can be seen by explicitly calculating $w_\phi$, which is equal to $w_{\rm de}$ today and is given by
\begin{equation}
w_\phi=\frac{\frac{1}{2}\dot{\phi}^2-V(\phi)}{\frac{1}{2}\dot{\phi}^2+V(\phi)}.
\label{eq:thescalareos}
\end{equation}
Using the potential in Eq.~\eqref{eq:potentialforsignflip} for $n=1$, together with $\dot\phi=H_c\phi'\simeq m_\phi\phi$, we obtain
\begin{equation}
w_\phi=
\frac{\frac{1}{2}m_\phi^2\phi^2-\Lambda_{\rm de}^4\left(1-\frac{\phi^2}{\tilde M^2}\right)}
{\frac{1}{2}m_\phi^2\phi^2+\Lambda_{\rm de}^4\left(1-\frac{\phi^2}{\tilde M^2}\right)},
\end{equation}
which, after substituting $m_\phi^2=2\Lambda_{\rm de}^4/\tilde M^2$ and inserting the growing solution, yields
\begin{equation}
w_\phi=-1+2\frac{\phi_i^2}{\tilde M^2}\exp\left(2\frac{m_\phi}{H_c}\Delta N\right).
\end{equation}
This expression allows us to place the following lower bound on $ w_{\rm de,0}\equiv w_{\phi,0}$:
\begin{equation}
    w_{\rm de,0}>-1+2\frac{\phi_i^2}{M^2}\left(\frac{2.3}{\alpha}\right)^{{2m_\phi}/{3H_c}}
\end{equation}
Therefore, obtaining $w_{\rm de,0}$ close to $-1$ today while having a large initial tachyonic mass requires a tuned value of $\phi_i/\tilde M$. This is because if the field varies too much, the kinetic and potential contributions quickly drive the system away from accelerated expansion and may also push the field outside of the EFT regime. The EFT constraint arises from the definitions of $V(\phi)$ and $m(\phi)$ in Eq.~\eqref{eq:potentialforsignflip}, where a higher-order truncation is assumed. Field displacements larger than $M$ or $\tilde M$ would therefore introduce higher-order corrections that invalidate this truncation. From the definition of $\alpha$ in Eq.~\eqref{eq:alpha}, we see that $M>\tilde M$, so the EFT limit is first reached when $\Delta\phi>\tilde M$. Using Eq.~\eqref{eq:phi(N) n=1}, this implies
\begin{equation}
\frac{\phi_i}{\tilde M}
\left[
\exp\left(\frac{m_\phi}{H_c}\Delta N\right)-1
\right]\approx\frac{\phi_i}{\tilde{M}}\left[\left(\frac{2.3}{\alpha}\right)^{{m_\phi}/{3H_c}}-1\right]<1.
\end{equation}
This can be avoided by adding a minimum in the bare potential about which the field eventually oscillates, but these oscillations would not themselves provide accelerated expansion, and so an additional cosmological constant would still be required. A similar issue appears in the opposite limit: if the field remains stationary for too long, the dark matter mass does not evolve sufficiently to produce an effective phantom crossing. The same qualitative conclusion can be derived for $n>1$, as we show in Appendix~\ref{app: n>1}.

\begin{figure}
    \centering
    \includegraphics[width=1\linewidth]{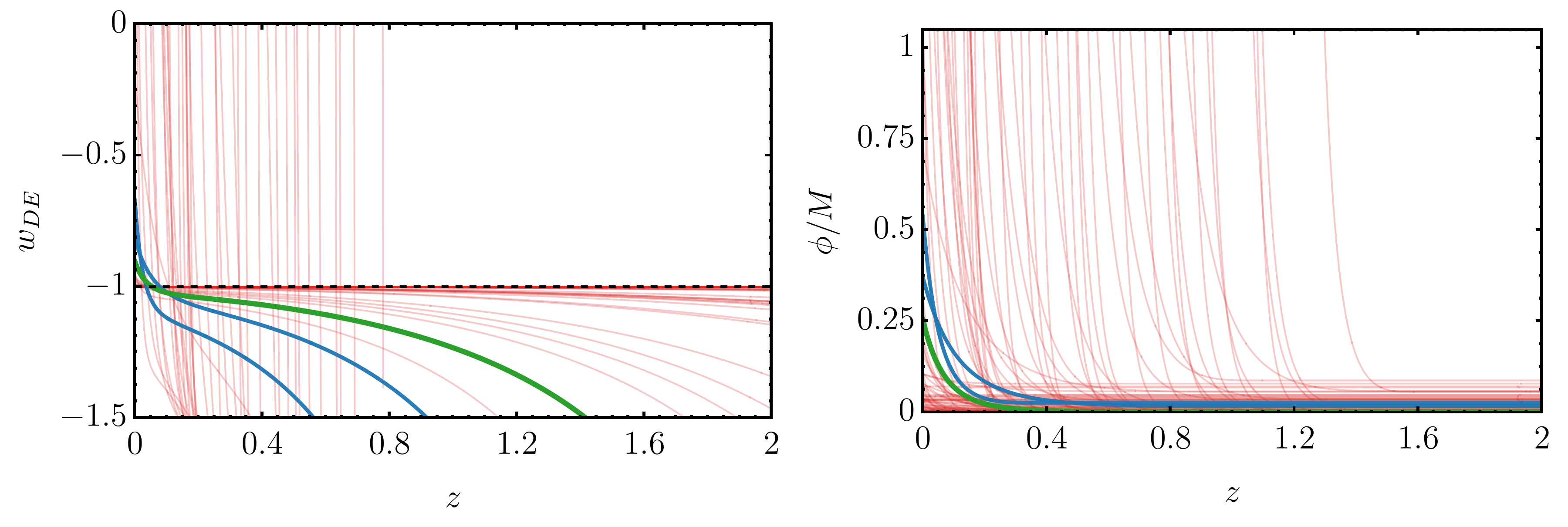}
    \caption{Evolution of the scalar field $\phi$ and the equation of state of the inferred dark-energy fluid, $w_{\rm de}$, for a random sample of $m_\phi^2/H_c^2$, $\phi_i/M$, and $\alpha$ using the full equations of motion from Eq.~\eqref{eq:signflip} with $n=1$. The scan uses log-uniform priors $\log_{10}(\phi_i/M)\in[-5,-1]$, $\log_{10}(m_\phi/H_c)\in[0,2]$, and $\alpha\in[0.05,1.5]$. Blue curves denote cases with $-0.9<w_{\rm de,0}<-0.5$, while red curves denote cases that fail this condition, either by overshooting or by remaining frozen too long and never crossing. In this scan, only $2/150$ points remain viable, although the sampled values are chosen near the dark-energy scale, as very large initial masses can drive $w_{\rm de}>0$ and push the system outside the EFT regime. The green curve shows a representative case with $m_\phi/H_c=9$, $\phi_i/M=10^{-3}$, and $\alpha=0.25$.}
    \label{fig:signflip}
\end{figure}

These analytical arguments are supported by a numerical evolution of the full equations of motion in Eq.~\eqref{eq:signflip}, shown in Figure~\ref{fig:signflip} for 150 random samples near the dark-energy scale. Viable phantom-crossing solutions can still be found, with $-0.9<w_{\rm de,0}<-0.5$, as illustrated by the two blue curves and the highlighted green curve. However, these choices are by necessity fine tuned. While we have focused on the case $\phi_i/\tilde M\ll1$, one could also take a smaller value of $m_\phi/H_c$, so that the field evolves more slowly after the transition. However, this weakens the main motivation for the sign-flip mechanism, as the transition should not only occur near the dark-energy scale, but should also make the field leave slow roll soon afterwards.

This mechanism therefore trades the coincidence of the field becoming dynamical near dark energy domination for a tuning of the field trajectory. Turning to more generic forms of $V(\phi)$ and allowing for $m_\phi<H_c$ may reduce these issues. This is the case where the field evolution is no longer controlled by a sign flip, but by a slow-roll exit at late times. We now investigate whether this possibility can provide a more robust way to produce the apparent phantom crossing.

\subsection{Slow-roll exit}

Here we allow for a generic bare potential, $V(\phi)$, and study the conditions that it must satisfy in order for the field to leave slow roll precisely when it starts to dominate the energy density. In this way, we do not address the cosmological constant problem itself, but rather the coincidence associated with the dark matter mass rescaling taking place near dark energy domination.

In this case, the dark matter mass interaction is already dynamically subdominant by the time $\phi$ starts rolling. The equation of motion can then be written as
\begin{align}\label{eq:phi eom slow roll}
{\phi}''+\left(3+\frac{H'}{H}\right){\phi}'\approx-\frac{V_{,\phi}(\phi)}{H^2},
\end{align}
where we have again used derivatives with respect to e-folds, $N=\log(a)$. In principle, one could make approximations similar to those used for the sign-flip mechanism, but they are less useful here because our goal is to study generic potentials. Instead, we take a semi-analytic approach, given the large literature existing on the evolution of quintessence-type models (see Ref.~\cite{Copeland:2006wr} for an in-depth review). Since we are mainly interested in the relation between $\Omega_\phi$ and $w_\phi$, it is convenient to rewrite the second-order equation as a first-order system,
\begin{equation}\label{eq: slowroll eqs}
\begin{split}
w_\phi'=&(1-w_\phi)\left[-3(1+w_\phi)+\lambda(\phi)\sqrt{3\Omega_\phi(1+w_\phi)}\right],\\
\Omega_\phi'=&-3w_\phi\Omega_\phi(1-\Omega_\phi),
\end{split}
\end{equation}
where we have set $\Omega_{\rm r}\approx0$, since here we are focused on late-time dark energy, and
\begin{equation}
\lambda(\phi)=-\Mpl\frac{V_{,\phi}}{V}.
\end{equation}
The field displacement is then obtained from
\begin{equation}
\phi'=\Mpl\sqrt{3(1+w_\phi)\Omega_\phi}.
\end{equation}
While this system can become complicated for general potentials, we can already see that the field remains close to $w_\phi=-1$ as long as $3\lambda^2\Omega_\phi\ll1$, which is the usual slow-roll regime. Once this condition is violated, the field leaves slow roll and $w_\phi$ increases as the right hand side of $w_\phi'$ becomes positive. The time at which this transition occurs is model dependent, because $\lambda(\phi)$ depends on $V(\phi)$, and therefore on the initial field configuration, $\phi_i$. However, there is one exception: if $\lambda$ is field independent, the evolution becomes insensitive to the initial value of $\phi$. This occurs for an exponential potential,
\begin{equation}
V(\phi)=V_0 \exp(-\tilde{\lambda}\phi/\Mpl).
\end{equation}
In this case, the field is initially frozen while $\Omega_\phi$ is small. As $\Omega_\phi$ grows, the slow-roll condition is eventually violated, approximately when $\Omega_\phi\sim1/3\tilde{\lambda}^2$, and $w_\phi$ starts evolving toward larger values. The final evolution is controlled by $\tilde{\lambda}$, since exponential potentials have a scalar-field dominated attractor with $\Omega_\phi=1$ for $\tilde{\lambda}<\sqrt{3}$~\cite{Copeland:1997et,Copeland:2006wr}, with equation of state
\begin{equation}
w_\phi=\frac{\tilde{\lambda}^2}{3}-1.
\end{equation}
The exponential potential therefore reduces the coincidence associated with having a varying field today, because the departure from slow roll is tied directly to the growth of $\Omega_\phi$, independently of the initial value of $\phi_i$. Once dark energy starts dominating, the exponential potential induces motion in the scalar field, which then drives the apparent phantom crossing through the time variation of the dark matter mass. This behavior is not generic to any potential, given the variation of $\lambda$ as $\phi$ evolves. In some cases, $\lambda(\phi)$ may become large long before $\Omega_\phi$ is close to one, causing the mass rescaling to occur during the dark ages. In other cases, $\lambda(\phi)$ may remain too small for the field to move appreciably at late times. However, this does not mean that only the exponential potential gives the desired behavior, as one can identify initial conditions for which a generic potential mimics the exponential case.

\begin{figure}
    \centering
    \includegraphics[width=1\linewidth]{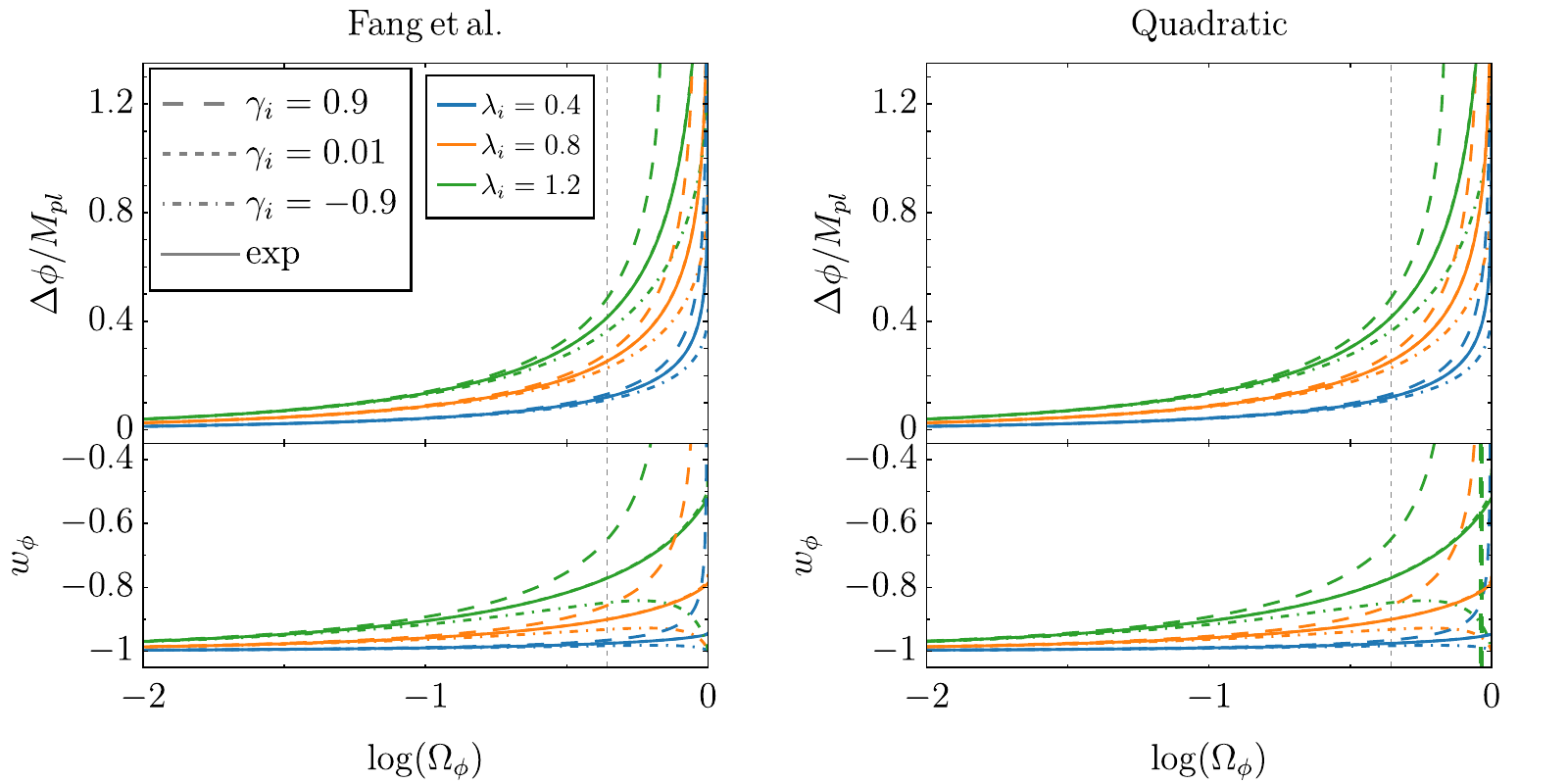}
    \caption{Evolution of $\Delta\phi$ and $w_\phi$ for the Fang et al. and quadratic potentials, with initial conditions and parameters chosen to give different values of $\lambda_i$ and $\gamma_i$. Smaller values of $|\gamma_i|$ lead to a closer tracking of the exponential-potential evolution, shown by the solid line. The horizontal axis is the scalar energy fraction, illustrating that the exponential potential becomes dynamical when $\Omega_\phi$ starts becoming relevant, independently of the initial field value. The vertical dashed line marks the present-day value of $\log(\Omega_\phi)$, which equals $\log(\Omega_{\rm de})$ by definition. }
    \label{fig:FangQuad}
\end{figure}

A useful way to identify these initial conditions is to require the potential to mimic the exponential evolution over the relevant field range. To do this, we track the variation of $\lambda$ and require it to remain small. Using $\lambda=-\Mpl V_{,\phi}/V$, we find
\begin{equation}
\frac{d\lambda}{dN}
=
-\Mpl\left(\frac{V_{,\phi\phi}V-V_{,\phi}^2}{V^2}\right)\phi'
=
\gamma(\phi)\frac{\phi'}{\Mpl},
\end{equation}
where for convenience we define 
\begin{equation}
\gamma(\phi)=-\Mpl^2\left(\frac{V_{,\phi\phi}V-V_{,\phi}^2}{V^2}\right),
\end{equation}
which is nothing but the sensitivity of $\lambda$ to variations of the field. Therefore, for a generic potential to behave similarly to an exponential potential with a given $\tilde{\lambda}$, its parameters must satisfy $\lambda_i\equiv \lambda(\phi_i)=\tilde{\lambda}$ and $|\gamma(\phi)|\ll1$ over the field range relevant for the late-time evolution. We illustrate this with two examples:

\textbf{The Fang et al. model.}~\cite{Fang:2008fw} This corresponds to the closest extension of the exponential potential and is obtained by allowing for quadratic terms in the exponent,
\begin{align}
V(\phi)&=V_0 \exp\left(-\beta\frac{\phi^2}{\Mpl^2}\right).
\end{align}
Using the definitions of $\lambda(\phi)$ and $\gamma(\phi)$ we obtain
\begin{align}
\lambda(\phi)&=2\beta\frac{\phi}{\Mpl},  &
\gamma(\phi)&=2\beta.
\end{align}
From these, we see that mimicking the exponential behavior fixes $\phi_i$ to provide a given value  $\lambda(\phi_i)$, and requires $|\beta|\ll1$, so that $\lambda$ varies slowly. We show the dependence of the evolution on these parameters in Figure~\ref{fig:FangQuad}.

\textbf{Quadratic potential.} A second useful example is the quadratic potential. This is easily motivated by a $\mathbb{Z}_2$ symmetry, and particularly interesting since such theories do not show long range fifth forces. The potential is given by
\begin{align}
V(\phi)&=V_0\left(1+\beta \frac{\phi^2}{\Mpl^2}\right), 
\end{align}
with
\begin{align}
\lambda(\phi)&=-\frac{2\beta\Mpl\phi}{\Mpl^2+\beta\phi^2}, &
\gamma(\phi)&=\frac{2\beta\Mpl^2(\beta\phi^2-\Mpl^2)}{(\Mpl^2+\beta\phi^2)^2}.
\end{align}
To find the parameters that mimic an exponential potential, we first impose $\lambda_i=\tilde{\lambda}$, which gives
\begin{equation}
\beta=-\frac{\tilde{\lambda}\Mpl^2}{\phi_i(2\Mpl+\tilde{\lambda}\phi_i)}.
\end{equation}
Substituting this into $\gamma(\phi_i)$, we find
\begin{equation}
\phi_i=\frac{\tilde{\lambda}\Mpl}{\gamma_i-\tilde{\lambda}^2}.
\end{equation}
Thus, for a fixed $\tilde{\lambda}$, the initial condition is determined by how slowly $\lambda$ is allowed to vary. For $\gamma_i>\tilde{\lambda}^2$, the required initial value of $\phi_i$ is positive, while for $\gamma_i<\tilde{\lambda}^2$ it is negative. The sign of $\beta$ is fixed by the corresponding signs of $\phi_i$ and $\tilde{\lambda}$ that fix the initial condition condition $\lambda_i=\tilde{\lambda}$. This illustrates the advantage of the method: without assuming a specific sign for the coupling, both branches can be identified as potentials that approximately reproduce the exponential evolution over the relevant field range. For the branch in which $\beta<0$, we impose $1+\beta\phi^2/\Mpl^2>0$ so that the dark energy contribution to the expansion remains positive. The evolution of this system for different values of $\tilde{\lambda}$ and $\gamma_i$ is shown in Figure~\ref{fig:FangQuad}.

\begin{figure}
    \centering
    \includegraphics[width=1\linewidth]{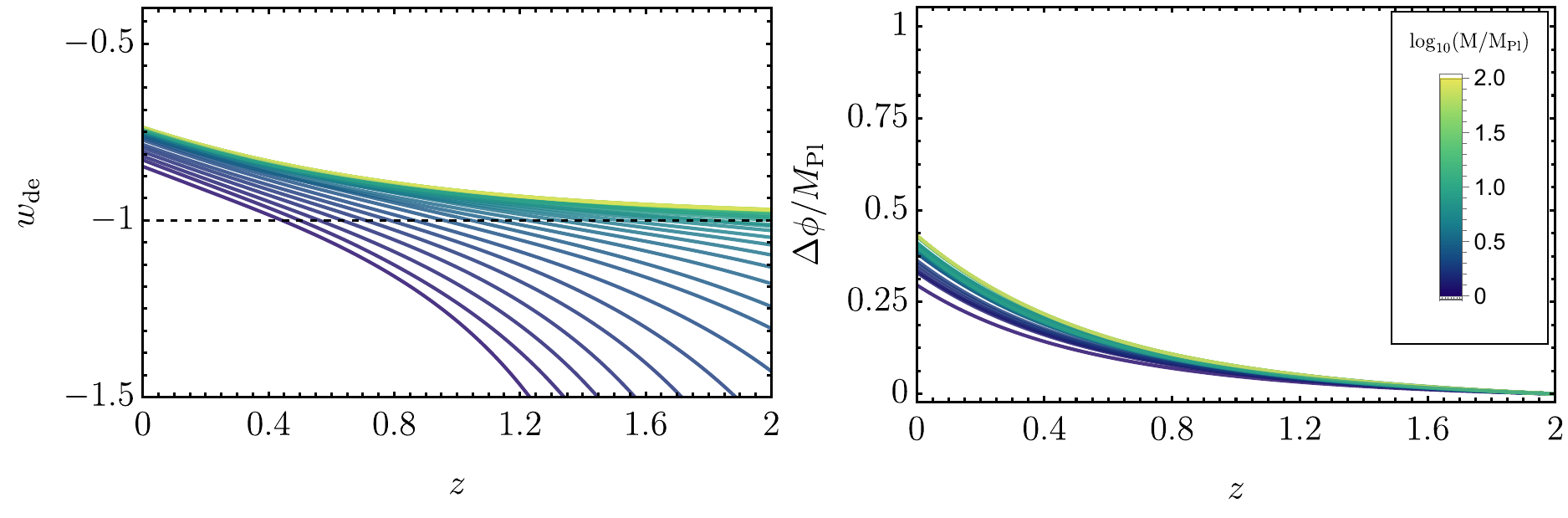}
    \caption{Evolution of $\phi$ and $w_{\rm de}$ for an exponential potential with $\lambda=1$ and different choices of $M>M_{\rm Pl}$, in the regime where the field evolution is approximately decoupled from $m(\phi)$. The field begins to evolve only near the present epoch through ordinary slow roll, effectively independently of $\phi_i$. All choices approach $w_{\rm de,0}\simeq -0.7$, as expected from the exponential-potential fixed point, while the preceding phantom crossing depends on the coupling strength. For $M\gtrsim 100M_{\rm Pl}$, the interaction is too weak to produce a crossing, indicating that this effect is controlled most efficiently by scales $M\sim M_{\rm Pl}$.}
    \label{fig:Exp_phantom}
\end{figure}

The impact of this behavior on the dark matter mass, and therefore on the evolution of the effective dark-energy fluid, depends on the form of $m(\phi)$. Following Section~\ref{sec:DEMeq}, we take
\begin{equation}
m(\phi)=m_0\left(1+\frac{\phi^{2n}}{M^{2n}}\right).
\end{equation}
For the field to evolve primarily under the effects of $V(\phi)$, with its dynamics effectively decoupled from the mass variation, the curvature induced by the coupling must remain sub-Hubble, $m_{\rm eff}^2/H_0^2<1$. At the same time, producing a phantom crossing requires the dark matter mass to be sensitive to field excursions of order $\Delta\phi\sim 0.5M_{\rm Pl}$, as in the exponential potential evolution shown in Figure~\ref{fig:FangQuad}. This points to interaction scales $M\gtrsim M_{\rm Pl}$, as illustrated in Figure~\ref{fig:Exp_phantom}, where the full equations of motion show that for $1<M/M_{\rm Pl}<100$ one obtains a well-behaved phantom crossing without tuning the scale of $V(\phi)$. Because the exponential potential becomes dynamical naturally when it begins to dominate the energy density, the crossing is then controlled mainly by the value of $M$. However, notice that one can achieve a similar effect without such large values of $M$ if only a sector of dark matter is coupled to $\phi$, making $(1-r)$ smaller. This behavior is much less tuned than in the sign-flip case and can be easily generalized to other potentials by finding the corresponding $\lambda_i$ and $\gamma_i$.

\begin{figure}
    \centering
    \includegraphics[width=1\linewidth]{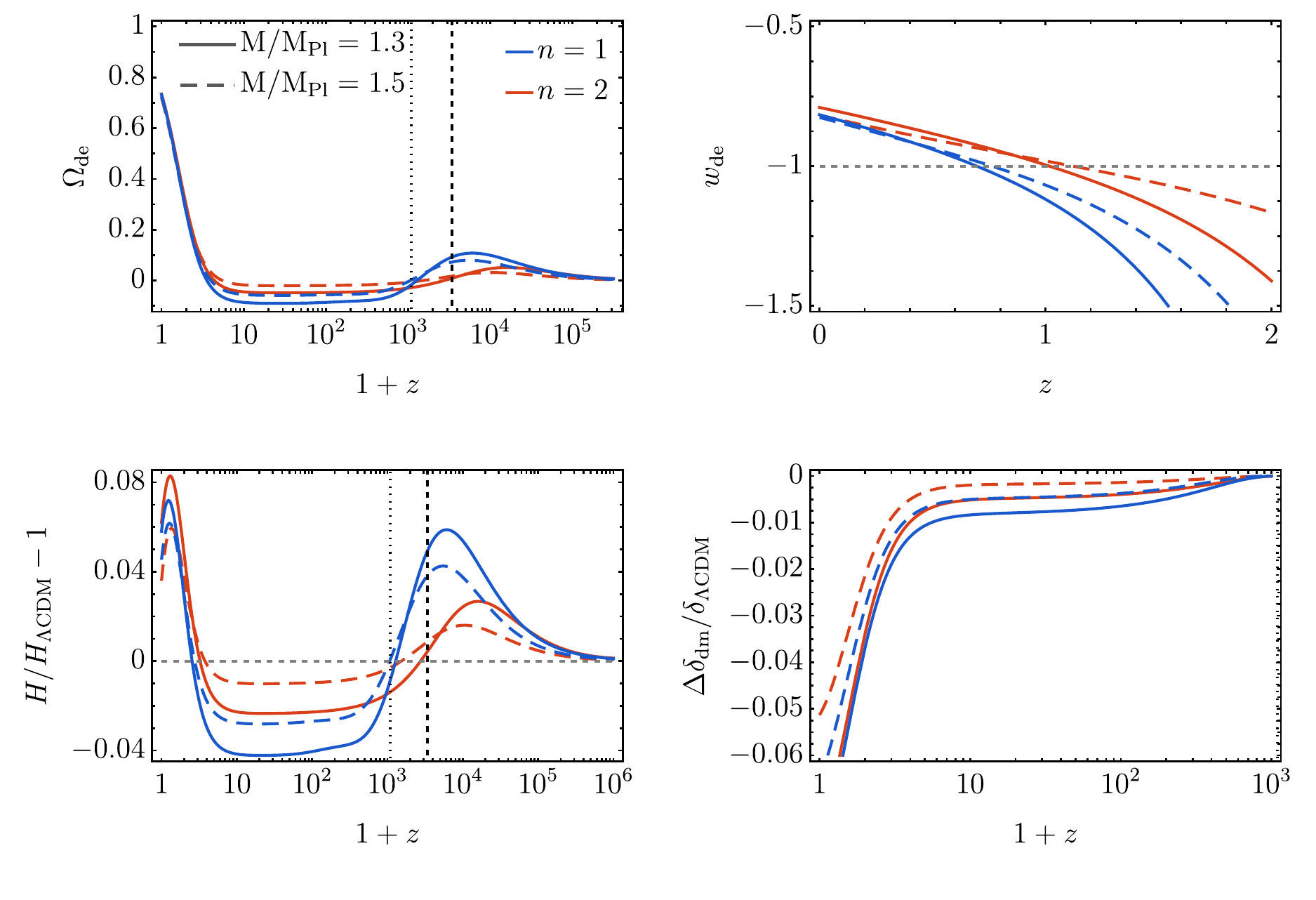}
    \caption{Scalar-field evolution for an exponential potential and a choice of $m(\phi)$ with $n=1$ and $n=2$ couplings between dark matter and a quintessence field $\phi$ from $z=10^6$ until present day. For each $n$, we show two choices of the mass scale $M$ and $\phi_i$. On the bottom plots we show the relative difference between the model and the inferred $\Lambda$CDM evolution of the Hubble parameter (left) and dark matter structure growth (right). We can see now that for $M\approx1.5\Mpl$ one can achieve an $\mathcal{O}(10\%)$ peak at matter-radiation equality (dashed vertical line) that turns off before recombination (dotted vertical line), a late time phantom crossing around dark energy-matter equality and a suppressed dark matter clustering during the dark ages.}
    \label{fig:all}
\end{figure}

To conclude this section, we also contrast these results with those of Section~\ref{sec:MReq}, unifying the early- and late-time effects on cosmology within the same model. This comparison is shown in Figure~\ref{fig:all} for $n=1$ and $n=2$. As we can see, the same values of $M$ that lead to a well-behaved phantom crossing at late times make the mass coupling become important near matter-radiation equality and produce an effective dark-energy peak with amplitude $\mathcal{O}(10\%)$. The late-time crossing is largely insensitive to $\phi_i$ for the exponential potentials considered here, while the early-time peak depends directly on the initial displacement, since it appears only when $m(\phi_i)>m(\phi_0)$. Such a displacement may be difficult to motivate in isolation, but it can arise naturally if $\phi$ also has interactions with the Standard Model~\cite{Burrage:2026pei}. 

\begin{figure}
    \centering
    \includegraphics[width=1\linewidth]{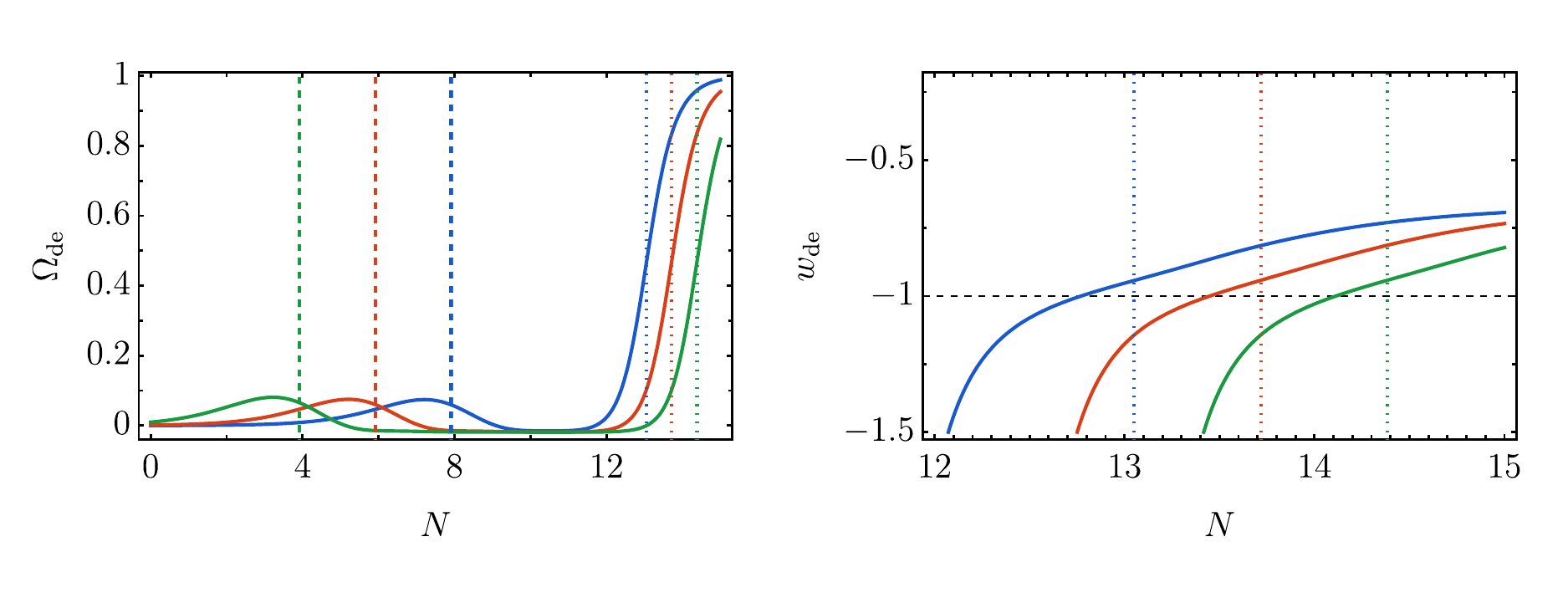}
    \caption{Scalar-field evolution for an exponential potential and a mass function $m(\phi)$ with $n=1$, $M=1.5\Mpl$, and $\phi_i=0.8M$. For each colored curve, we choose different energy scales for matter-radiation equality, shown by the dashed vertical lines, and matter-dark-energy equality, shown by the dotted vertical lines. The separation between these two scales is chosen to be smaller than in our Universe for the blue curve, equal to that of our Universe for the red curve, and larger for the green curve. This illustrates that for fixed parameters and initial conditions, the early dark-energy peak always occurs around the corresponding epoch of matter-radiation equality, while the phantom crossing point similarly tracks the matter-dark-energy scale.}
    \label{fig: timigs}
\end{figure}

It is important to point out that the fact that the same range of $M$ controls both effects is not tied to the particular numerical values of the matter-radiation equality or dark-energy scales. Rather, the interaction is sourced by the relevant background abundance, $\Omega_{\rm m}$ at early times and $\Omega_{\rm de}$ at late times, as we illustrate in Figure~\ref{fig: timigs} by changing these two scales while keeping a fixed choice of model parameters and initial conditions. This provides a natural way to soften the coincidence of the field being relevant at these two scales, while also supplying the ingredients needed to reduce the sound horizon at early times and generate an effective phantom crossing today. Moreover, because the dark matter mass changes in time, the same mechanism can reduce dark matter clustering.

\section{Discussion and outlook}\label{sec:Conclusion}

In this paper, we have studied the connection between dark-sector interactions and cosmological tensions using analytical arguments. We focused on a class of models in which the dark matter mass depends on a scalar field, $m(\phi)$, and asked under what conditions a single field can produce two separate effects over the cosmological history. The first is an energy injection around matter-radiation equality, and the second is an apparent phantom crossing at late times. These two regimes are directly connected to two of the main observational motivations for physics beyond $\Lambda$CDM, namely the Hubble tension and the DESI preference for evolving dark energy.

We first studied the early-time dynamics around matter-radiation equality, where the coupling to dark matter can temporarily displace the scalar field and generate an EDE-like contribution to the expansion rate while reducing the $S_8$ tension. We showed that the time of the peak is controlled mainly by the local curvature of $m(\phi)$ around the initial field value, while the amplitude is controlled by the ratio between the initial and present-day dark matter masses. This separation provides a useful analytic way to understand how the early-time energy injection can be adjusted without fully specifying the late-time scalar potential.

We then turned to the late-time evolution, where the bare potential $V(\phi)$ moves the field toward its present-day value, $\phi_0$. Since the dark matter mass must remain approximately constant between recombination and the onset of dark energy domination, the key question is why the field starts moving only at low redshift, for which we considered two possibilities. The sign-flip mechanism naturally ties the onset of the instability to matter-dark energy equality, but trades this coincidence for a tuning of the field trajectory, which is highly sensitive to the initial displacement and tachyonic mass. The slow-roll exit scenario provides a more robust possibility, in which the field remains frozen until $V(\phi)$ becomes dynamically relevant. In particular, exponential potentials play a special role because their constant slope parameter ($\lambda=-\Mpl V_{,\phi}/V$) ties the departure from slow roll directly to the growth of its energy fraction, making the late-time motion insensitive to the initial value of $\phi$. This behavior can also be generalized to any potential for which $\lambda(\phi)$ varies slowly over the relevant field range, providing a simple analytic criterion for constructing more general potentials with the same late-time behavior.

Putting these results together, we find that the same scalar-dependent dark matter mass can generate an early-time energy injection when the dark matter abundance becomes important near matter-radiation equality, and a late-time phantom mirage when the bare potential drives the field at low redshift. The main reduction in tuning for this model is therefore a reduction in timing coincidences. In the effective dark energy description, both effects arise from the mismatch between the physical dark matter density and the density inferred under the assumption of constant-mass cold dark matter. The early peak is controlled mainly by the structure of $m(\phi)$, while the late phantom regime depends on the interplay between the $V(\phi)$, and the resulting change in $m(\phi)$. This full behavior is shown in Figure~\ref{fig:all}.

In conclusion, our aim has been to understand minimal mass-varying dark matter models analytically and to develop theoretical tools for identifying when they work. This provides a useful framework for isolating the regions of parameter space in which a simultaneous impact on early- and late-time cosmological tensions can be expected. A natural continuation of this work is to confront the same model simultaneously with the full set of cosmological observations.

\section*{Acknowledgments}
The authors thank Ed Copeland, Justin Khoury and Meng-Xiang Lin for useful discussions. SSM is supported by funds provided by the Center for Particle Cosmology at the University of Pennsylvania. The work of MT is supported in part by DOE (HEP) Award No. DE-SC0013528.

\appendix
\section{Sign-flip with n>1}\label{app: n>1}
We can extend the fine tuning arguments from Section~\ref{sec:DEMeq signflip} to the case $n>1$. The conservative equation of motion after the sign flip is
\begin{equation}
\label{eq: tachyonic approx n greater 1}
\phi''+3\phi'
-2n\frac{\Lambda_{\rm de}^4}{\tilde M^{2n}}
\frac{\phi^{2n-1}}{H_c^2}=0.
\end{equation}
Unlike the $n=1$ case, this equation is nonlinear, and the instability is not controlled by a constant tachyonic mass. Instead, the relevant local mass depends on the initial displacement,
\begin{equation}
m_\phi^2(\phi_i)=2n(2n-1)\frac{\Lambda_{\rm de}^4}{\tilde M^{2n}}\phi_i^{2n-2}.
\end{equation}
Once the acceleration term becomes subdominant, Eq.~\eqref{eq: tachyonic approx n greater 1} gives the terminal-velocity approximation
\begin{equation}
\phi'=\frac{1}{3(2n-1)}\frac{m_\phi^2(\phi_i)}{H_c^2}\frac{\phi^{2n-1}}{\phi_i^{2n-2}}.
\end{equation}
Solving this equation yields
\begin{equation}\label{eq: terminal velocity n greater 1}
\phi(N)=\phi_i\left[1-\frac{2n-2}{3(2n-1)}\frac{m_\phi^2(\phi_i)}{H_c^2}\Delta N\right]^{-1/(2n-2)}.
\end{equation}
Therefore, although the growth is no longer purely exponential, the evolution is still highly sensitive to the ratio $m_\phi(\phi_i)/H_c$ and to the initial displacement $\phi_i$. The apparent divergence of Eq.~\eqref{eq: terminal velocity n greater 1} should not be interpreted as a physical singularity, but as a breakdown of the terminal-velocity approximation. Before this point is reached, the kinetic energy becomes important, $w_\phi\to1$, and the system enters a kination regime. However, such an evolution is already incompatible with the desired late-time behavior, since the field no longer behaves as an evolving dark-energy component close to $w_\phi\simeq -1$.

As in the $n=1$ case, the tension can be seen directly from the scalar equation of state~(\ref{eq:thescalareos}).

Using $H=H_c$, the terminal-velocity approximation, and the potential~(\ref{eq:potentialforsignflip})
we obtain
\begin{equation}
w_\phi=\frac{\frac{n}{9(2n-1)}\frac{m_\phi^2(\phi_i)}{H_c^2}
\frac{\phi^{4n-2}}{\phi_i^{2n-2}\tilde M^{2n}}-\left(1-\frac{\phi^{2n}}{\tilde M^{2n}}\right)}{\frac{n}{9(2n-1)}\frac{m_\phi^2(\phi_i)}{H_c^2}\frac{\phi^{4n-2}}{\phi_i^{2n-2}\tilde M^{2n}}+\left(1-\frac{\phi^{2n}}{\tilde M^{2n}}\right)}.
\end{equation}
Substituting Eq.~\eqref{eq: terminal velocity n greater 1}, this becomes
\begin{equation}
w_\phi=\frac{\frac{\phi_i^{2n}}{\tilde M^{2n}}\left[\frac{n}{9(2n-1)}\frac{m_\phi^2(\phi_i)}{H_c^2}A^{-\frac{4n-2}{2n-2}}+A^{-\frac{2n}{2n-2}}\right]-1}{\frac{\phi_i^{2n}}{\tilde M^{2n}}\left[\frac{n}{9(2n-1)}\frac{m_\phi^2(\phi_i)}{H_c^2}A^{-\frac{4n-2}{2n-2}}-A^{-\frac{2n}{2n-2}}\right]+1}.
\end{equation}
with
5
\begin{equation}
A=1-\frac{2n-2}{3(2n-1)}\frac{m_\phi^2(\phi_i)}{H_c^2}\Delta N,
\end{equation}
where $A>0$ from Eq.~\eqref{eq: terminal velocity n greater 1}. This expression shows the same qualitative tension found for $n=1$. If $\phi_i/\tilde M$ is very small, then $w_\phi\simeq -1$, but the field remains too close to the maximum and the dark matter mass does not change enough to produce an observable phantom mirage. If instead $m_\phi(\phi_i)/H_c$ is large enough to generate a sizable displacement, the powers of $A^{-1}$ quickly enhance the kinetic contribution and drive the field away from $w_\phi\simeq -1$, while also risking loss of EFT control. Thus the $n>1$ case does not remove the tuning found in the quadratic case. It only replaces the exponential sensitivity of the $n=1$ solution by a nonlinear sensitivity to the initial displacement and the local tachyonic mass.

\bibliographystyle{IEEEtran}
\bibliography{refs}

@article{Burrage:2026pei,
    author = "Burrage, Clare and Sevillano Mu{\~n}oz, Sergio",
    title = "{Misalignment from kicks: the impact of particle interactions on ultra-light dark matter}",
    eprint = "2603.28881",
    archivePrefix = "arXiv",
    primaryClass = "hep-ph",
    month = "3",
    year = "2026"
}

@article{Khoury:2025txd,
    author = "Khoury, Justin and Lin, Meng-Xiang and Trodden, Mark",
    title = "{Apparent w{\ensuremath{<}}-1 and a Lower S8 from Dark Axion and Dark Baryons Interactions}",
    eprint = "2503.16415",
    archivePrefix = "arXiv",
    primaryClass = "astro-ph.CO",
    doi = "10.1103/w4qb-plk8",
    journal = "Phys. Rev. Lett.",
    volume = "135",
    number = "18",
    pages = "181001",
    year = "2025"
}

@article{Carroll:2004hc,
    author = "Carroll, Sean M. and De Felice, Antonio and Trodden, Mark",
    title = "{Can we be tricked into thinking that w is less than -1?}",
    eprint = "astro-ph/0408081",
    archivePrefix = "arXiv",
    doi = "10.1103/PhysRevD.71.023525",
    journal = "Phys. Rev. D",
    volume = "71",
    pages = "023525",
    year = "2005"
}

@article{Liu:2023rvo,
    author = "Liu, Gang and Gao, Jiaze and Han, Yufen and Mu, Yuhao and Xu, Lixin",
    title = "{Coupled dark sector models and cosmological tensions}",
    eprint = "2312.01410",
    archivePrefix = "arXiv",
    primaryClass = "astro-ph.CO",
    doi = "10.1103/PhysRevD.109.103531",
    journal = "Phys. Rev. D",
    volume = "109",
    number = "10",
    pages = "103531",
    year = "2024"
}

@article{Kamionkowski:2024axz,
    author = "Kamionkowski, Marc and Mathur, Anubhav",
    title = "{Thermocoupled early dark energy}",
    eprint = "2411.09747",
    archivePrefix = "arXiv",
    primaryClass = "hep-ph",
    doi = "10.1103/PhysRevD.111.063551",
    journal = "Phys. Rev. D",
    volume = "111",
    number = "6",
    pages = "063551",
    year = "2025"
}

@article{Lin:2022phm,
    author = "Lin, Meng-Xiang and McDonough, Evan and Hill, J. Colin and Hu, Wayne",
    title = "{Dark matter trigger for early dark energy coincidence}",
    eprint = "2212.08098",
    archivePrefix = "arXiv",
    primaryClass = "astro-ph.CO",
    doi = "10.1103/PhysRevD.107.103523",
    journal = "Phys. Rev. D",
    volume = "107",
    number = "10",
    pages = "103523",
    year = "2023"
}

@article{Liu:2025bss,
    author = "Liu, Rayne and Zhu, Yijie and Hu, Wayne and Miranda, Vivian",
    title = "{Phantom mirage from axion dark energy}",
    eprint = "2510.14957",
    archivePrefix = "arXiv",
    primaryClass = "astro-ph.CO",
    doi = "10.1103/3s1m-9zpc",
    journal = "Phys. Rev. D",
    volume = "113",
    number = "8",
    pages = "083506",
    year = "2026"
}

@article{Wang:2026wrk,
    author = "Wang, Jia-Qi and Cai, Rong-Gen and Guo, Zong-Kuan and Li, Yun-He and Wang, Shao-Jiang and Zhang, Xin",
    title = "{Non-minimally coupled quintessence with sign-switching interaction}",
    eprint = "2604.02204",
    archivePrefix = "arXiv",
    primaryClass = "astro-ph.CO",
    month = "4",
    year = "2026"
}

@article{Naidoo:2026umv,
    author = "Naidoo, Krishna and Hallam, James and Baker, Tessa and Sirera, Sergi",
    title = "{Constraints on Horndeski Gravity with Phantom Crossing}",
    eprint = "2606.20794",
    archivePrefix = "arXiv",
    primaryClass = "astro-ph.CO",
    month = "6",
    year = "2026"
}

@article{SanchezLopez:2025uzw,
    author = "S{\'a}nchez L{\'o}pez, Samuel and Karam, Alexandros and Hazra, Dhiraj Kumar",
    title = "{Non-Minimally Coupled Quintessence in Light of DESI}",
    eprint = "2510.14941",
    archivePrefix = "arXiv",
    primaryClass = "astro-ph.CO",
    month = "10",
    year = "2025"
}

@article{Copeland:2023zqz,
    author = "Copeland, Edmund J. and Moss, Adam and Sevillano Mu{\~n}oz, Sergio and White, Jade M. M.",
    title = "{Scaling solutions as Early Dark Energy resolutions to the Hubble tension}",
    eprint = "2309.15295",
    archivePrefix = "arXiv",
    primaryClass = "astro-ph.CO",
    doi = "10.1088/1475-7516/2024/05/078",
    journal = "JCAP",
    volume = "05",
    pages = "078",
    year = "2024"
}

@article{Verde:2019ivm,
    author = "Verde, L. and Treu, T. and Riess, A. G.",
    title = "{Tensions between the Early and the Late Universe}",
    eprint = "1907.10625",
    archivePrefix = "arXiv",
    primaryClass = "astro-ph.CO",
    doi = "10.1038/s41550-019-0902-0",
    journal = "Nature Astron.",
    volume = "3",
    pages = "891",
    year = "2019"
}

@article{DiValentino:2020zio,
    author = "Di Valentino, Eleonora and others",
    title = "{Snowmass2021 - Letter of interest cosmology intertwined II: The hubble constant tension}",
    eprint = "2008.11284",
    archivePrefix = "arXiv",
    primaryClass = "astro-ph.CO",
    reportNumber = "FERMILAB-PUB-21-590-PPD",
    doi = "10.1016/j.astropartphys.2021.102605",
    journal = "Astropart. Phys.",
    volume = "131",
    pages = "102605",
    year = "2021"
}

@article{DiValentino:2021izs,
    author = "Di Valentino, Eleonora and Mena, Olga and Pan, Supriya and Visinelli, Luca and Yang, Weiqiang and Melchiorri, Alessandro and Mota, David F. and Riess, Adam G. and Silk, Joseph",
    title = "{In the realm of the Hubble tension{\textemdash}a review of solutions}",
    eprint = "2103.01183",
    archivePrefix = "arXiv",
    primaryClass = "astro-ph.CO",
    reportNumber = "IPPP/20/108",
    doi = "10.1088/1361-6382/ac086d",
    journal = "Class. Quant. Grav.",
    volume = "38",
    number = "15",
    pages = "153001",
    year = "2021"
}

@article{Schoneberg:2021qvd,
    author = {Sch{\"o}neberg, Nils and Franco Abell{\'a}n, Guillermo and P{\'e}rez S{\'a}nchez, Andrea and Witte, Samuel J. and Poulin, Vivian and Lesgourgues, Julien},
    title = "{The H0 Olympics: A fair ranking of proposed models}",
    eprint = "2107.10291",
    archivePrefix = "arXiv",
    primaryClass = "astro-ph.CO",
    doi = "10.1016/j.physrep.2022.07.001",
    journal = "Phys. Rept.",
    volume = "984",
    pages = "1--55",
    year = "2022"
}

@article{Kamionkowski:2022pkx,
    author = "Kamionkowski, Marc and Riess, Adam G.",
    title = "{The Hubble Tension and Early Dark Energy}",
    eprint = "2211.04492",
    archivePrefix = "arXiv",
    primaryClass = "astro-ph.CO",
    doi = "10.1146/annurev-nucl-111422-024107",
    journal = "Ann. Rev. Nucl. Part. Sci.",
    volume = "73",
    pages = "153--180",
    year = "2023"
}

@article{CosmoVerseNetwork:2025alb,
    author = "Di Valentino, Eleonora and others",
    collaboration = "CosmoVerse Network",
    title = "{The CosmoVerse White Paper: Addressing observational tensions in cosmology with systematics and fundamental physics}",
    eprint = "2504.01669",
    archivePrefix = "arXiv",
    primaryClass = "astro-ph.CO",
    doi = "10.1016/j.dark.2025.101965",
    journal = "Phys. Dark Univ.",
    volume = "49",
    pages = "101965",
    year = "2025"
}

@article{Planck:2018vyg,
    author = "Aghanim, N. and others",
    collaboration = "Planck",
    title = "{Planck 2018 results. VI. Cosmological parameters}",
    eprint = "1807.06209",
    archivePrefix = "arXiv",
    primaryClass = "astro-ph.CO",
    doi = "10.1051/0004-6361/201833910",
    journal = "Astron. Astrophys.",
    volume = "641",
    pages = "A6",
    year = "2020",
    note = "[Erratum: Astron.Astrophys. 652, C4 (2021)]"
}

@article{ACT:2023dou,
    author = "Qu, Frank J. and others",
    collaboration = "ACT",
    title = "{The Atacama Cosmology Telescope: A Measurement of the DR6 CMB Lensing Power Spectrum and Its Implications for Structure Growth}",
    eprint = "2304.05202",
    archivePrefix = "arXiv",
    primaryClass = "astro-ph.CO",
    reportNumber = "FERMILAB-PUB-23-237-PPD, FERMILAB-PUB-23-237-PPD",
    doi = "10.3847/1538-4357/acfe06",
    journal = "Astrophys. J.",
    volume = "962",
    number = "2",
    pages = "112",
    year = "2024"
}

@article{ACT:2023kun,
    author = "Madhavacheril, Mathew S. and others",
    collaboration = "ACT",
    title = "{The Atacama Cosmology Telescope: DR6 Gravitational Lensing Map and Cosmological Parameters}",
    eprint = "2304.05203",
    archivePrefix = "arXiv",
    primaryClass = "astro-ph.CO",
    reportNumber = "FERMILAB-PUB-23-206-PPD",
    doi = "10.3847/1538-4357/acff5f",
    journal = "Astrophys. J.",
    volume = "962",
    number = "2",
    pages = "113",
    year = "2024"
}

@article{Riess:2021jrx,
    author = "Riess, Adam G. and others",
    title = "{A Comprehensive Measurement of the Local Value of the Hubble Constant with 1 km/s/Mpc Uncertainty from the Hubble Space Telescope and the SH0ES Team}",
    eprint = "2112.04510",
    archivePrefix = "arXiv",
    primaryClass = "astro-ph.CO",
    doi = "10.3847/2041-8213/ac5c5b",
    journal = "Astrophys. J. Lett.",
    volume = "934",
    number = "1",
    pages = "L7",
    year = "2022"
}

@article{Poulin:2025nfb,
    author = "Poulin, Vivian and Smith, Tristan L. and Calder{\'o}n, Rodrigo and Simon, Th{\'e}o",
    title = "{Impact of ACT DR6 and DESI DR2 for early dark energy and the Hubble tension}",
    eprint = "2505.08051",
    archivePrefix = "arXiv",
    primaryClass = "astro-ph.CO",
    doi = "10.1103/bx25-1g5d",
    journal = "Phys. Rev. D",
    volume = "113",
    number = "6",
    pages = "063519",
    year = "2026"
}

@article{Poulin:2018cxd,
    author = "Poulin, Vivian and Smith, Tristan L. and Karwal, Tanvi and Kamionkowski, Marc",
    title = "{Early Dark Energy Can Resolve The Hubble Tension}",
    eprint = "1811.04083",
    archivePrefix = "arXiv",
    primaryClass = "astro-ph.CO",
    doi = "10.1103/PhysRevLett.122.221301",
    journal = "Phys. Rev. Lett.",
    volume = "122",
    number = "22",
    pages = "221301",
    year = "2019"
}

@article{Alexander:2019rsc,
    author = "Alexander, Stephon and McDonough, Evan",
    title = "{Axion-Dilaton Destabilization and the Hubble Tension}",
    eprint = "1904.08912",
    archivePrefix = "arXiv",
    primaryClass = "astro-ph.CO",
    doi = "10.1016/j.physletb.2019.134830",
    journal = "Phys. Lett. B",
    volume = "797",
    pages = "134830",
    year = "2019"
}

@article{Smith:2019ihp,
    author = "Smith, Tristan L. and Poulin, Vivian and Amin, Mustafa A.",
    title = "{Oscillating scalar fields and the Hubble tension: a resolution with novel signatures}",
    eprint = "1908.06995",
    archivePrefix = "arXiv",
    primaryClass = "astro-ph.CO",
    doi = "10.1103/PhysRevD.101.063523",
    journal = "Phys. Rev. D",
    volume = "101",
    number = "6",
    pages = "063523",
    year = "2020"
}

@article{Sakstein:2019fmf,
    author = "Sakstein, Jeremy and Trodden, Mark",
    title = "{Early Dark Energy from Massive Neutrinos as a Natural Resolution of the Hubble Tension}",
    eprint = "1911.11760",
    archivePrefix = "arXiv",
    primaryClass = "astro-ph.CO",
    doi = "10.1103/PhysRevLett.124.161301",
    journal = "Phys. Rev. Lett.",
    volume = "124",
    number = "16",
    pages = "161301",
    year = "2020"
}

@article{DESI:2024mwx,
    author = "Adame, A. G. and others",
    collaboration = "DESI",
    title = "{DESI 2024 VI: cosmological constraints from the measurements of baryon acoustic oscillations}",
    eprint = "2404.03002",
    archivePrefix = "arXiv",
    primaryClass = "astro-ph.CO",
    reportNumber = "FERMILAB-PUB-24-0154-PPD",
    doi = "10.1088/1475-7516/2025/02/021",
    journal = "JCAP",
    volume = "02",
    pages = "021",
    year = "2025"
}

@article{DESI:2025zgx,
    author = "Abdul Karim, M. and others",
    collaboration = "DESI",
    title = "{DESI DR2 results. II. Measurements of baryon acoustic oscillations and cosmological constraints}",
    eprint = "2503.14738",
    archivePrefix = "arXiv",
    primaryClass = "astro-ph.CO",
    reportNumber = "FERMILAB-PUB-25-0169-PPD",
    doi = "10.1103/tr6y-kpc6",
    journal = "Phys. Rev. D",
    volume = "112",
    number = "8",
    pages = "083515",
    year = "2025"
}

@article{Pan:2023mie,
    author = "Pan, Supriya and Yang, Weiqiang",
    title = "{On the interacting dark energy scenarios $-$ the case for Hubble constant tension}",
    eprint = "2310.07260",
    archivePrefix = "arXiv",
    primaryClass = "astro-ph.CO",
    doi = "10.1007/978-981-99-0177-7_29",
    month = "10",
    year = "2023"
}

@article{DESI:2025fii,
    author = "Lodha, K. and others",
    collaboration = "DESI",
    title = "{Extended dark energy analysis using DESI DR2 BAO measurements}",
    eprint = "2503.14743",
    archivePrefix = "arXiv",
    primaryClass = "astro-ph.CO",
    reportNumber = "FERMILAB-PUB-25-0164-PPD",
    doi = "10.1103/w4c6-1r5j",
    journal = "Phys. Rev. D",
    volume = "112",
    number = "8",
    pages = "083511",
    year = "2025"
}

@article{Bernui:2023byc,
    author = "Bernui, Armando and Di Valentino, Eleonora and Giar{\`e}, William and Kumar, Suresh and Nunes, Rafael C.",
    title = "{Exploring the H0 tension and the evidence for dark sector interactions from 2D BAO measurements}",
    eprint = "2301.06097",
    archivePrefix = "arXiv",
    primaryClass = "astro-ph.CO",
    doi = "10.1103/PhysRevD.107.103531",
    journal = "Phys. Rev. D",
    volume = "107",
    number = "10",
    pages = "103531",
    year = "2023"
}

@article{Figueruelo:2026eis,
    author = "Figueruelo, David and van der Westhuizen, Marcel and Abebe, Amare and Di Valentino, Eleonora",
    title = "{Late-time background constraints on linear and non-linear interacting dark energy after DESI DR2}",
    eprint = "2601.03122",
    archivePrefix = "arXiv",
    primaryClass = "astro-ph.CO",
    doi = "10.1016/j.dark.2026.102238",
    journal = "Phys. Dark Univ.",
    volume = "52",
    pages = "102238",
    year = "2026"
}

@article{Giare:2024smz,
    author = "Giar{\`e}, William and Sabogal, Miguel A. and Nunes, Rafael C. and Di Valentino, Eleonora",
    title = "{Interacting Dark Energy after DESI Baryon Acoustic Oscillation Measurements}",
    eprint = "2404.15232",
    archivePrefix = "arXiv",
    primaryClass = "astro-ph.CO",
    doi = "10.1103/PhysRevLett.133.251003",
    journal = "Phys. Rev. Lett.",
    volume = "133",
    number = "25",
    pages = "251003",
    year = "2024"
}

@article{DiValentino:2017iww,
    author = "Di Valentino, Eleonora and Melchiorri, Alessandro and Mena, Olga",
    title = "{Can interacting dark energy solve the $H_0$ tension?}",
    eprint = "1704.08342",
    archivePrefix = "arXiv",
    primaryClass = "astro-ph.CO",
    doi = "10.1103/PhysRevD.96.043503",
    journal = "Phys. Rev. D",
    volume = "96",
    number = "4",
    pages = "043503",
    year = "2017"
}

@article{DiValentino:2019jae,
    author = "Di Valentino, Eleonora and Melchiorri, Alessandro and Mena, Olga and Vagnozzi, Sunny",
    title = "{Nonminimal dark sector physics and cosmological tensions}",
    eprint = "1910.09853",
    archivePrefix = "arXiv",
    primaryClass = "astro-ph.CO",
    doi = "10.1103/PhysRevD.101.063502",
    journal = "Phys. Rev. D",
    volume = "101",
    number = "6",
    pages = "063502",
    year = "2020"
}

@article{Yashiki:2025loj,
    author = "Yashiki, Mai",
    title = "{Toward a simultaneous resolution of the H0 and S8 tensions: Early dark energy and an interacting dark sector model}",
    eprint = "2505.23382",
    archivePrefix = "arXiv",
    primaryClass = "astro-ph.CO",
    doi = "10.1103/qw1d-mdrz",
    journal = "Phys. Rev. D",
    volume = "112",
    number = "6",
    pages = "063517",
    year = "2025"
}

@article{Yao:2025wlx,
    author = "Yao, Zhibang and Ye, Gen and Silvestri, Alessandra",
    title = "{A general model for dark energy crossing the phantom divide}",
    eprint = "2508.01378",
    archivePrefix = "arXiv",
    primaryClass = "gr-qc",
    doi = "10.1088/1475-7516/2025/10/078",
    journal = "JCAP",
    volume = "10",
    pages = "078",
    year = "2025"
}

@article{Antusch:2026ldp,
    author = "Antusch, Stefan and King, Stephen F. and Wang, Xin",
    title = "{Coupled Dark Energy and Dark Matter for DESI: An Effective Guide to the Phantom Divide}",
    eprint = "2604.08449",
    archivePrefix = "arXiv",
    primaryClass = "astro-ph.CO",
    month = "4",
    year = "2026"
}

@article{Giare:2026tyk,
    author = "Giar{\`e}, William and Sakstein, Jeremy",
    title = "{Unifying Early and Late Dark Energy: Dynamical Requirements and Obstructions}",
    eprint = "2605.26116",
    archivePrefix = "arXiv",
    primaryClass = "astro-ph.CO",
    month = "5",
    year = "2026"
}

@article{Sohail:2024oki,
    author = "Sohail, Sk. and Alam, Sonej and Akthar, Shiriny and Hossain, Md. Wali",
    title = "{Quintessential early dark energy}",
    eprint = "2408.03229",
    archivePrefix = "arXiv",
    primaryClass = "astro-ph.CO",
    doi = "10.1016/j.dark.2025.101948",
    journal = "Phys. Dark Univ.",
    volume = "48",
    pages = "101948",
    year = "2025"
}

@article{Adil:2022hkj,
    author = "Adil, Arsalan and Albrecht, Andreas and Knox, Lloyd",
    title = "{Quintessential cosmological tensions}",
    eprint = "2207.10235",
    archivePrefix = "arXiv",
    primaryClass = "astro-ph.CO",
    doi = "10.1103/PhysRevD.107.063521",
    journal = "Phys. Rev. D",
    volume = "107",
    number = "6",
    pages = "063521",
    year = "2023"
}

@article{Ramadan:2023ivw,
    author = "Ramadan, Omar F. and Karwal, Tanvi and Sakstein, Jeremy",
    title = "{Attractive proposal for resolving the Hubble tension: Dynamical attractors that unify early and late dark energy}",
    eprint = "2309.08082",
    archivePrefix = "arXiv",
    primaryClass = "astro-ph.CO",
    doi = "10.1103/PhysRevD.109.063525",
    journal = "Phys. Rev. D",
    volume = "109",
    number = "6",
    pages = "063525",
    year = "2024"
}

@article{Turner:1983he,
    author = "Turner, Michael S.",
    title = "{Coherent Scalar Field Oscillations in an Expanding Universe}",
    reportNumber = "EFI-83-29-CHICAGO",
    doi = "10.1103/PhysRevD.28.1243",
    journal = "Phys. Rev. D",
    volume = "28",
    pages = "1243",
    year = "1983"
}

@article{Copeland:2006wr,
    author = "Copeland, Edmund J. and Sami, M. and Tsujikawa, Shinji",
    title = "{Dynamics of dark energy}",
    eprint = "hep-th/0603057",
    archivePrefix = "arXiv",
    doi = "10.1142/S021827180600942X",
    journal = "Int. J. Mod. Phys. D",
    volume = "15",
    pages = "1753--1936",
    year = "2006"
}

@article{Copeland:1997et,
    author = "Copeland, Edmund J. and Liddle, Andrew R and Wands, David",
    title = "{Exponential potentials and cosmological scaling solutions}",
    eprint = "gr-qc/9711068",
    archivePrefix = "arXiv",
    reportNumber = "SUSX-TH-97-022, SUSSEX-AST-97-11-1, PU-RCG-97-20",
    doi = "10.1103/PhysRevD.57.4686",
    journal = "Phys. Rev. D",
    volume = "57",
    pages = "4686--4690",
    year = "1998"
}

@article{Cai:2021weh,
    author = "Cai, Rong-Gen and Guo, Zong-Kuan and Wang, Shao-Jiang and Yu, Wang-Wei and Zhou, Yong",
    title = "{No-go guide for the Hubble tension: Late-time solutions}",
    eprint = "2107.13286",
    archivePrefix = "arXiv",
    primaryClass = "astro-ph.CO",
    doi = "10.1103/PhysRevD.105.L021301",
    journal = "Phys. Rev. D",
    volume = "105",
    number = "2",
    pages = "L021301",
    year = "2022"
}

@article{Bansal:2026axl,
    author = "Bansal, Prakhar and Huterer, Dragan",
    title = "{On the Difficulties with Late-Time Solutions for the Hubble Tension}",
    eprint = "2602.06293",
    archivePrefix = "arXiv",
    primaryClass = "astro-ph.CO",
    month = "2",
    year = "2026"
}

@article{Garcia-Garcia:2026nzy,
    author = "Garc{\'\i}a-Garc{\'\i}a, Carlos and Ferreira, Pedro G. and Wolf, William J.",
    title = "{The Status of Single Scalar Field Dark Energy}",
    eprint = "2607.07777",
    archivePrefix = "arXiv",
    primaryClass = "astro-ph.CO",
    month = "7",
    year = "2026"
}

@article{Wolf:2024eph,
    author = "Wolf, William J. and Garc{\'\i}a-Garc{\'\i}a, Carlos and Bartlett, Deaglan J. and Ferreira, Pedro G.",
    title = "{Scant evidence for thawing quintessence}",
    eprint = "2408.17318",
    archivePrefix = "arXiv",
    primaryClass = "astro-ph.CO",
    doi = "10.1103/PhysRevD.110.083528",
    journal = "Phys. Rev. D",
    volume = "110",
    number = "8",
    pages = "083528",
    year = "2024"
}

@article{Calderon:2026hbr,
    author = "Calderon, Rodrigo and Linder, Eric V.",
    title = "{Charging Across the Phantom Divide with Modified Gravity}",
    eprint = "2605.26259",
    archivePrefix = "arXiv",
    primaryClass = "gr-qc",
    month = "5",
    year = "2026"
}

@article{Caldwell:2025inn,
    author = "Caldwell, Robert R. and Linder, Eric V.",
    title = "{Null impact of the null energy condition in current cosmology}",
    eprint = "2511.07526",
    archivePrefix = "arXiv",
    primaryClass = "astro-ph.CO",
    doi = "10.1088/1475-7516/2026/05/008",
    journal = "JCAP",
    volume = "05",
    pages = "008",
    year = "2026"
}

@article{Fang:2008fw,
    author = "Fang, Wei and Li, Ying and Zhang, Kai and Lu, Hui-Qing",
    title = "{Exact Analysis of Scaling and Dominant Attractors Beyond the Exponential Potential}",
    eprint = "0810.4193",
    archivePrefix = "arXiv",
    primaryClass = "hep-th",
    doi = "10.1088/0264-9381/26/15/155005",
    journal = "Class. Quant. Grav.",
    volume = "26",
    pages = "155005",
    year = "2009"
}

@inproceedings{Teixeira:2026yjd,
    author = "Teixeira, Elsa M.",
    title = "{Interactions in the dark sector: intrinsic entropy couplings}",
    booktitle = "{60th Rencontres de Moriond on Cosmology}: {Moriond Cosmology 2026}",
    eprint = "2606.16574",
    archivePrefix = "arXiv",
    primaryClass = "astro-ph.CO",
    month = "6",
    year = "2026"
}

@article{Jensko:2026taf,
    author = "Jensko, Erik and Teixeira, Elsa M. and Poulin, Vivian",
    title = "{Interacting dark sector from intrinsic entropy couplings}",
    eprint = "2603.10622",
    archivePrefix = "arXiv",
    primaryClass = "astro-ph.CO",
    month = "3",
    year = "2026"
}

@article{Teixeira:2024qmw,
    author = "Teixeira, Elsa M. and Poulot, Gaspard and van de Bruck, Carsten and Di Valentino, Eleonora and Poulin, Vivian",
    title = "{Alleviating cosmological tensions with a hybrid dark sector}",
    eprint = "2412.14139",
    archivePrefix = "arXiv",
    primaryClass = "astro-ph.CO",
    doi = "10.1103/9lf2-33zf",
    journal = "Phys. Rev. D",
    volume = "113",
    number = "2",
    pages = "023514",
    year = "2026"
}

@article{vandeBruck:2022xbk,
    author = "van de Bruck, Carsten and Poulot, Gaspard and Teixeira, Elsa M.",
    title = "{Scalar field dark matter and dark energy: a hybrid model for the dark sector}",
    eprint = "2211.13653",
    archivePrefix = "arXiv",
    primaryClass = "hep-th",
    doi = "10.1088/1475-7516/2023/07/019",
    journal = "JCAP",
    volume = "07",
    pages = "019",
    year = "2023"
}

@article{Poulot:2024sex,
    author = "Poulot, Gaspard and Teixeira, Elsa M. and van de Bruck, Carsten and Nunes, Nelson J.",
    title = "{Scalar field dark matter with time-varying equation of state}",
    eprint = "2404.10524",
    archivePrefix = "arXiv",
    primaryClass = "astro-ph.CO",
    doi = "10.1088/1475-7516/2025/11/005",
    journal = "JCAP",
    volume = "11",
    pages = "005",
    year = "2025"
}

@article{Smith:2025uaq,
    author = "Smith, Adam and Mylova, Maria and van de Bruck, Carsten and Burgess, C. P. and Di Valentino, Eleonora",
    title = "{The Serendipitous Axiodilaton: A Self-Consistent Recombination-Era Solution to the Hubble Tension}",
    eprint = "2512.13544",
    archivePrefix = "arXiv",
    primaryClass = "astro-ph.CO",
    month = "12",
    year = "2025"
}

@article{Pettorino:2013ia,
    author = "Pettorino, Valeria and Amendola, Luca and Wetterich, Christof",
    title = "{How early is early dark energy?}",
    eprint = "1301.5279",
    archivePrefix = "arXiv",
    primaryClass = "astro-ph.CO",
    doi = "10.1103/PhysRevD.87.083009",
    journal = "Phys. Rev. D",
    volume = "87",
    pages = "083009",
    year = "2013"
}

@article{Gogoi:2020qif,
    author = "Gogoi, Antareep and Sharma, Ravi Kumar and Chanda, Prolay and Das, Subinoy",
    title = "{Early Mass-varying Neutrino Dark Energy: Nugget Formation and Hubble Anomaly}",
    eprint = "2005.11889",
    archivePrefix = "arXiv",
    primaryClass = "astro-ph.CO",
    reportNumber = "2021 ApJ 915 132",
    doi = "10.3847/1538-4357/abfe5b",
    journal = "Astrophys. J.",
    volume = "915",
    number = "2",
    pages = "132",
    year = "2021"
}

@article{Kamionkowski:2014zda,
    author = "Kamionkowski, Marc and Pradler, Josef and Walker, Devin G. E.",
    title = "{Dark energy from the string axiverse}",
    eprint = "1409.0549",
    archivePrefix = "arXiv",
    primaryClass = "hep-ph",
    reportNumber = "SLAC-PUB-16085",
    doi = "10.1103/PhysRevLett.113.251302",
    journal = "Phys. Rev. Lett.",
    volume = "113",
    number = "25",
    pages = "251302",
    year = "2014"
}

@article{Cortes:2024lgw,
    author = "Cort{\^e}s, Marina and Liddle, Andrew R.",
    title = "{Interpreting DESI's evidence for evolving dark energy}",
    eprint = "2404.08056",
    archivePrefix = "arXiv",
    primaryClass = "astro-ph.CO",
    doi = "10.1088/1475-7516/2024/12/007",
    journal = "JCAP",
    volume = "12",
    pages = "007",
    year = "2024"
}

@article{Shlivko:2024llw,
    author = "Shlivko, David and Steinhardt, Paul J.",
    title = "{Assessing observational constraints on dark energy}",
    eprint = "2405.03933",
    archivePrefix = "arXiv",
    primaryClass = "astro-ph.CO",
    doi = "10.1016/j.physletb.2024.138826",
    journal = "Phys. Lett. B",
    volume = "855",
    pages = "138826",
    year = "2024"
}

@article{Shlivko:2025fgv,
    author = "Shlivko, David and Steinhardt, Paul J. and Steinhardt, Charles L.",
    title = "{Optimal parameterizations for observational constraints on thawing dark energy}",
    eprint = "2504.02028",
    archivePrefix = "arXiv",
    primaryClass = "astro-ph.CO",
    doi = "10.1088/1475-7516/2025/06/054",
    month = "4",
    year = "2025"
}

@article{Shlivko:2026jxa,
    author = "Shlivko, David and Poulin, Vivian",
    title = "{Phantom-Crossing Dark Energy and the $\Omega_m$ Tug-of-War}",
    eprint = "2603.22406",
    archivePrefix = "arXiv",
    primaryClass = "astro-ph.CO",
    month = "3",
    year = "2026"
}

@article{Delaunay:2026jto,
    author = "Delaunay, C{\'e}dric and Greljo, Admir",
    title = "{Natural Phantom Dark Energy from a $\mathbb{Z}_N$--Axion}",
    eprint = "2607.06774",
    archivePrefix = "arXiv",
    primaryClass = "hep-ph",
    month = "7",
    year = "2026"
}

@article{Amendola:1999er,
    author = "Amendola, Luca",
    title = "{Coupled quintessence}",
    eprint = "astro-ph/9908023",
    archivePrefix = "arXiv",
    doi = "10.1103/PhysRevD.62.043511",
    journal = "Phys. Rev. D",
    volume = "62",
    pages = "043511",
    year = "2000"
}

@article{Doran:2006kp,
    author = "Doran, Michael and Robbers, Georg",
    title = "{Early dark energy cosmologies}",
    eprint = "astro-ph/0601544",
    archivePrefix = "arXiv",
    reportNumber = "HD-THEP-06-01",
    doi = "10.1088/1475-7516/2006/06/026",
    journal = "JCAP",
    volume = "06",
    pages = "026",
    year = "2006"
}

@article{Amendola:2006dg,
    author = "Amendola, Luca and Camargo Campos, Gabriela and Rosenfeld, Rogerio",
    title = "{Consequences of dark matter-dark energy interaction on cosmological parameters derived from SNIa data}",
    eprint = "astro-ph/0610806",
    archivePrefix = "arXiv",
    reportNumber = "IFT-P-035-2006",
    doi = "10.1103/PhysRevD.75.083506",
    journal = "Phys. Rev. D",
    volume = "75",
    pages = "083506",
    year = "2007"
}

@article{Heymans:2020gsg,
    author = "Heymans, Catherine and others",
    title = "{KiDS-1000 Cosmology: Multi-probe weak gravitational lensing and spectroscopic galaxy clustering constraints}",
    eprint = "2007.15632",
    archivePrefix = "arXiv",
    primaryClass = "astro-ph.CO",
    doi = "10.1051/0004-6361/202039063",
    journal = "Astron. Astrophys.",
    volume = "646",
    pages = "A140",
    year = "2021"
}

@article{DES:2021wwk,
    author = "Abbott, T. M. C. and others",
    collaboration = "DES",
    title = "{Dark Energy Survey Year 3 results: Cosmological constraints from galaxy clustering and weak lensing}",
    eprint = "2105.13549",
    archivePrefix = "arXiv",
    primaryClass = "astro-ph.CO",
    reportNumber = "FERMILAB-PUB-21-221-AE, DES-2020-0617",
    doi = "10.1103/PhysRevD.105.023520",
    journal = "Phys. Rev. D",
    volume = "105",
    number = "2",
    pages = "023520",
    year = "2022"
}

@article{Caldwell:2005tm,
    author = "Caldwell, R. R. and Linder, Eric V.",
    title = "{The Limits of quintessence}",
    eprint = "astro-ph/0505494",
    archivePrefix = "arXiv",
    doi = "10.1103/PhysRevLett.95.141301",
    journal = "Phys. Rev. Lett.",
    volume = "95",
    pages = "141301",
    year = "2005"
}

@article{Sekiguchi:2020teg,
    author = "Sekiguchi, Toyokazu and Takahashi, Tomo",
    title = "{Early recombination as a solution to the $H_0$ tension}",
    eprint = "2007.03381",
    archivePrefix = "arXiv",
    primaryClass = "astro-ph.CO",
    reportNumber = "KEK-TH-2238",
    doi = "10.1103/PhysRevD.103.083507",
    journal = "Phys. Rev. D",
    volume = "103",
    number = "8",
    pages = "083507",
    year = "2021"
}

@article{Thomas:2016iav,
    author = "Thomas, Daniel B. and Kopp, Michael and Skordis, Constantinos",
    title = "{Constraining the Properties of Dark Matter with Observations of the Cosmic Microwave Background}",
    eprint = "1601.05097",
    archivePrefix = "arXiv",
    primaryClass = "astro-ph.CO",
    doi = "10.3847/0004-637X/830/2/155",
    journal = "Astrophys. J.",
    volume = "830",
    number = "2",
    pages = "155",
    year = "2016"
}

@article{Lin:2019qug,
    author = "Lin, Meng-Xiang and Benevento, Giampaolo and Hu, Wayne and Raveri, Marco",
    title = "{Acoustic Dark Energy: Potential Conversion of the Hubble Tension}",
    eprint = "1905.12618",
    archivePrefix = "arXiv",
    primaryClass = "astro-ph.CO",
    doi = "10.1103/PhysRevD.100.063542",
    journal = "Phys. Rev. D",
    volume = "100",
    number = "6",
    pages = "063542",
    year = "2019"
}

@article{Huey:2004qv,
    author = "Huey, Greg and Wandelt, Benjamin D.",
    title = "{Interacting quintessence. The Coincidence problem and cosmic acceleration}",
    eprint = "astro-ph/0407196",
    archivePrefix = "arXiv",
    doi = "10.1103/PhysRevD.74.023519",
    journal = "Phys. Rev. D",
    volume = "74",
    pages = "023519",
    year = "2006"
}

@article{Das:2005yj,
    author = "Das, Subinoy and Corasaniti, Pier Stefano and Khoury, Justin",
    title = "{Super-acceleration as signature of dark sector interaction}",
    eprint = "astro-ph/0510628",
    archivePrefix = "arXiv",
    doi = "10.1103/PhysRevD.73.083509",
    journal = "Phys. Rev. D",
    volume = "73",
    pages = "083509",
    year = "2006"
}

@article{Brax:2004qh,
    author = "Brax, Philippe and van de Bruck, Carsten and Davis, Anne-Christine and Khoury, Justin and Weltman, Amanda",
    title = "{Detecting dark energy in orbit: The cosmological chameleon}",
    eprint = "astro-ph/0408415",
    archivePrefix = "arXiv",
    doi = "10.1103/PhysRevD.70.123518",
    journal = "Phys. Rev. D",
    volume = "70",
    pages = "123518",
    year = "2004"
}

@article{Smith:2025icl,
    author = {Smith, Adam and {\"O}z{\"u}lker, Emre and Di Valentino, Eleonora and van de Bruck, Carsten},
    title = "{Dynamical Dark Energy Meets Varying Electron Mass: Implications for Phantom Crossing and the Hubble Constant}",
    eprint = "2510.21931",
    archivePrefix = "arXiv",
    primaryClass = "astro-ph.CO",
    month = "10",
    year = "2025"
}

@article{Lee:2025pzo,
    author = "Lee, Dong Ha and Yang, Weiqiang and Di Valentino, Eleonora and Pan, Supriya and van de Bruck, Carsten",
    title = "{Shape of dark energy: Constraining its evolution with a general parametrization}",
    eprint = "2507.11432",
    archivePrefix = "arXiv",
    primaryClass = "astro-ph.CO",
    doi = "10.1103/z7y2-yvhg",
    journal = "Phys. Rev. D",
    volume = "113",
    number = "6",
    pages = "063554",
    year = "2026"
}

@article{Zhai:2025hfi,
    author = "Zhai, Yuejia and de Cesare, Marco and van de Bruck, Carsten and Di Valentino, Eleonora and Wilson-Ewing, Edward",
    title = "{A low-redshift preference for an interacting dark energy model}",
    eprint = "2503.15659",
    archivePrefix = "arXiv",
    primaryClass = "astro-ph.CO",
    doi = "10.1088/1475-7516/2025/11/010",
    journal = "JCAP",
    volume = "11",
    pages = "010",
    year = "2025"
}

@article{Smith:2025grk,
    author = "Smith, Adam and Brax, Philippe and van de Bruck, Carsten and Burgess, C. P. and Davis, Anne-Christine",
    title = "{Screened axio-dilaton cosmology: novel forms of early dark energy}",
    eprint = "2505.05450",
    archivePrefix = "arXiv",
    primaryClass = "hep-th",
    doi = "10.1140/epjc/s10052-025-14735-4",
    journal = "Eur. Phys. J. C",
    volume = "85",
    number = "9",
    pages = "1062",
    year = "2025"
}

@article{Smith:2024ibv,
    author = "Smith, Adam and Mylova, Maria and Brax, Philippe and van de Bruck, Carsten and Burgess, C. P. and Davis, Anne-Christine",
    title = "{A Minimal Axio-dilaton Dark Sector}",
    eprint = "2410.11099",
    archivePrefix = "arXiv",
    primaryClass = "hep-th",
    doi = "10.1088/1475-7516/2025/07/023",
    month = "10",
    year = "2024"
}

@article{Smith:2024ayu,
    author = "Smith, Adam and Mylova, Maria and Brax, Philippe and van de Bruck, Carsten and Burgess, C. P. and Davis, Anne-Christine",
    title = "{CMB implications of multi-field axio-dilaton cosmology}",
    eprint = "2408.10820",
    archivePrefix = "arXiv",
    primaryClass = "hep-th",
    doi = "10.1088/1475-7516/2024/12/058",
    journal = "JCAP",
    volume = "12",
    pages = "058",
    year = "2024"
}

@article{Zhai:2023yny,
    author = "Zhai, Yuejia and Giar{\`e}, William and van de Bruck, Carsten and Di Valentino, Eleonora and Mena, Olga and Nunes, Rafael C.",
    title = "{A consistent view of interacting dark energy from multiple CMB probes}",
    eprint = "2303.08201",
    archivePrefix = "arXiv",
    primaryClass = "astro-ph.CO",
    doi = "10.1088/1475-7516/2023/07/032",
    journal = "JCAP",
    volume = "07",
    pages = "032",
    year = "2023"
}

@article{Scherer:2025esj,
    author = "Scherer, Mateus and Sabogal, Miguel A. and Nunes, Rafael C. and De Felice, Antonio",
    title = "{Challenging the {\ensuremath{\Lambda}}CDM model: 5{\ensuremath{\sigma}} evidence for a dynamical dark energy late-time transition}",
    eprint = "2504.20664",
    archivePrefix = "arXiv",
    primaryClass = "astro-ph.CO",
    doi = "10.1103/n86r-sjgm",
    journal = "Phys. Rev. D",
    volume = "112",
    number = "4",
    pages = "043513",
    year = "2025"
}

@article{Ramadan:2024kmn,
    author = "Ramadan, Omar F. and Sakstein, Jeremy and Rubin, David",
    title = "{DESI constraints on exponential quintessence}",
    eprint = "2405.18747",
    archivePrefix = "arXiv",
    primaryClass = "astro-ph.CO",
    doi = "10.1103/PhysRevD.110.L041303",
    journal = "Phys. Rev. D",
    volume = "110",
    number = "4",
    pages = "L041303",
    year = "2024"
}

@article{Wang:2024dka,
    author = "Wang, Hao and Piao, Yun-Song",
    title = "{Dark energy in light of DESI DR1 and Hubble tension}",
    eprint = "2404.18579",
    archivePrefix = "arXiv",
    primaryClass = "astro-ph.CO",
    doi = "10.1016/j.physletb.2026.140180",
    journal = "Phys. Lett. B",
    volume = "873",
    pages = "140180",
    year = "2026"
}

@article{Ye:2024ywg,
    author = "Ye, Gen and Martinelli, Matteo and Hu, Bin and Silvestri, Alessandra",
    title = "{Hints of Nonminimally Coupled Gravity in DESI 2024 Baryon Acoustic Oscillation Measurements}",
    eprint = "2407.15832",
    archivePrefix = "arXiv",
    primaryClass = "astro-ph.CO",
    doi = "10.1103/PhysRevLett.134.181002",
    journal = "Phys. Rev. Lett.",
    volume = "134",
    number = "18",
    pages = "181002",
    year = "2025"
}

@article{Bhattacharya:2024hep,
    author = "Bhattacharya, Sukannya and Borghetto, Giulia and Malhotra, Ameek and Parameswaran, Susha and Tasinato, Gianmassimo and Zavala, Ivonne",
    title = "{Cosmological constraints on curved quintessence}",
    eprint = "2405.17396",
    archivePrefix = "arXiv",
    primaryClass = "astro-ph.CO",
    doi = "10.1088/1475-7516/2024/09/073",
    journal = "JCAP",
    volume = "09",
    pages = "073",
    year = "2024"
}

@article{Notari:2024rti,
    author = "Notari, Alessio and Redi, Michele and Tesi, Andrea",
    title = "{Consistent theories for the DESI dark energy fit}",
    eprint = "2406.08459",
    archivePrefix = "arXiv",
    primaryClass = "astro-ph.CO",
    doi = "10.1088/1475-7516/2024/11/025",
    journal = "JCAP",
    volume = "11",
    pages = "025",
    year = "2024"
}

@article{Orchard:2024bve,
    author = "Orchard, Lili and C{\'a}rdenas, V{\'\i}ctor H.",
    title = "{Probing dark energy evolution post-DESI 2024}",
    eprint = "2407.05579",
    archivePrefix = "arXiv",
    primaryClass = "astro-ph.CO",
    doi = "10.1016/j.dark.2024.101678",
    journal = "Phys. Dark Univ.",
    volume = "46",
    pages = "101678",
    year = "2024"
}

@article{Giare:2024oil,
    author = "Giar{\`e}, William",
    title = "{Dynamical dark energy beyond Planck? Constraints from multiple CMB probes, DESI BAO, and type-Ia supernovae}",
    eprint = "2409.17074",
    archivePrefix = "arXiv",
    primaryClass = "astro-ph.CO",
    doi = "10.1103/ss37-cxhn",
    journal = "Phys. Rev. D",
    volume = "112",
    number = "2",
    pages = "023508",
    year = "2025"
}

@article{Fikri:2024klc,
    author = "Fikri, Ramy and Elkhateeb, Esraa and Lashin, E. I. and El Hanafy, Waleed",
    title = "{A preference for dynamical phantom dark energy using one-parameter model with Planck, DESI DR1 BAO and SN data}",
    eprint = "2411.19362",
    archivePrefix = "arXiv",
    primaryClass = "astro-ph.CO",
    doi = "10.1016/j.aop.2025.170190",
    journal = "Annals Phys.",
    volume = "481",
    pages = "170190",
    year = "2025"
}

@article{Pourtsidou:2025sdd,
    author = "Pourtsidou, Alkistis",
    title = "{Exponential quintessence with momentum coupling to dark matter}",
    eprint = "2509.15091",
    archivePrefix = "arXiv",
    primaryClass = "astro-ph.CO",
    doi = "10.1088/1475-7516/2026/02/014",
    journal = "JCAP",
    volume = "02",
    pages = "014",
    year = "2026"
}

@article{Cai:2026swf,
    author = "Cai, Rong-Gen and Wang, Shao-Jiang",
    title = "{The Hubble tension: A decade review}",
    eprint = "2606.20434",
    archivePrefix = "arXiv",
    primaryClass = "astro-ph.CO",
    doi = "10.1088/1674-4527/ae842f",
    journal = "Res. Astron. Astrophys.",
    volume = "26",
    pages = "084011",
    year = "2026"
}

@article{Zheng:2024qzi,
    author = "Zheng, Jie and Qiang, Da-Chun and You, Zhi-Qiang",
    title = "{Cosmological constraints on dark energy models using DESI BAO 2024}",
    eprint = "2412.04830",
    archivePrefix = "arXiv",
    primaryClass = "astro-ph.CO",
    doi = "10.1088/1475-7516/2025/08/056",
    journal = "JCAP",
    volume = "08",
    pages = "056",
    year = "2025"
}

@article{Wolf:2025jlc,
    author = "Wolf, William J. and Garc{\'\i}a-Garc{\'\i}a, Carlos and Ferreira, Pedro G.",
    title = "{Robustness of dark energy phenomenology across different parameterizations}",
    eprint = "2502.04929",
    archivePrefix = "arXiv",
    primaryClass = "astro-ph.CO",
    doi = "10.1088/1475-7516/2025/05/034",
    journal = "JCAP",
    volume = "05",
    pages = "034",
    year = "2025"
}

@article{Shajib:2025tpd,
    author = "Shajib, Anowar J. and Frieman, Joshua A.",
    title = "{Scalar-field dark energy models: Current and forecast constraints}",
    eprint = "2502.06929",
    archivePrefix = "arXiv",
    primaryClass = "astro-ph.CO",
    doi = "10.1103/kjpb-r698",
    journal = "Phys. Rev. D",
    volume = "112",
    number = "6",
    pages = "063508",
    year = "2025"
}

@article{Roy:2025cxk,
    author = "Roy, Nandan and Chakrabarti, Soumya",
    title = "{Is phantom barrier crossing inevitable? A cosmographic analysis}",
    eprint = "2508.13740",
    archivePrefix = "arXiv",
    primaryClass = "astro-ph.CO",
    doi = "10.1140/epjc/s10052-026-15803-z",
    journal = "Eur. Phys. J. C",
    volume = "86",
    number = "5",
    pages = "586",
    year = "2026"
}

@article{Burgess:2021obw,
    author = "Burgess, C. P. and Dineen, Danielle and Quevedo, F.",
    title = "{Yoga Dark Energy: natural relaxation and other dark implications of a supersymmetric gravity sector}",
    eprint = "2111.07286",
    archivePrefix = "arXiv",
    primaryClass = "hep-th",
    reportNumber = "CERN-TH-2021-192",
    doi = "10.1088/1475-7516/2022/03/064",
    journal = "JCAP",
    volume = "03",
    number = "03",
    pages = "064",
    year = "2022"
}

@article{Jiang:2026cqh,
    author = "Jiang, Jun-Qian and Amin, Mustafa A. and Shafieloo, Arman",
    title = "{Late-Time Oscillating Quintessence in Light of DESI}",
    eprint = "2606.24221",
    archivePrefix = "arXiv",
    primaryClass = "astro-ph.CO",
    month = "6",
    year = "2026"
}

@article{Freese:2021rjq,
    author = "Freese, Katherine and Winkler, Martin Wolfgang",
    title = "{Chain early dark energy: A Proposal for solving the Hubble tension and explaining today{\textquoteright}s dark energy}",
    eprint = "2102.13655",
    archivePrefix = "arXiv",
    primaryClass = "astro-ph.CO",
    reportNumber = "UTTG-02-2021, NORDITA-2021-020",
    doi = "10.1103/PhysRevD.104.083533",
    journal = "Phys. Rev. D",
    volume = "104",
    number = "8",
    pages = "083533",
    year = "2021"
}

@article{Karwal:2021vpk,
    author = "Karwal, Tanvi and Raveri, Marco and Jain, Bhuvnesh and Khoury, Justin and Trodden, Mark",
    title = "{Chameleon early dark energy and the Hubble tension}",
    eprint = "2106.13290",
    archivePrefix = "arXiv",
    primaryClass = "astro-ph.CO",
    doi = "10.1103/PhysRevD.105.063535",
    journal = "Phys. Rev. D",
    volume = "105",
    number = "6",
    pages = "063535",
    year = "2022"
}

@article{Niedermann:2021ijp,
    author = "Niedermann, Florian and Sloth, Martin S.",
    title = "{Hot new early dark energy: Towards a unified dark sector of neutrinos, dark energy and dark matter}",
    eprint = "2112.00759",
    archivePrefix = "arXiv",
    primaryClass = "hep-ph",
    doi = "10.1016/j.physletb.2022.137555",
    journal = "Phys. Lett. B",
    volume = "835",
    pages = "137555",
    year = "2022"
}

@article{Toomey:2025yuy,
    author = "Toomey, Michael W. and Hughes, Ellie and Ivanov, Mikhail M. and Sullivan, James M.",
    title = "{Kinetic Mixing and the Phantom Illusion: Axion-Dilaton Quintessence in Light of DESI DR2}",
    eprint = "2511.23463",
    archivePrefix = "arXiv",
    primaryClass = "astro-ph.CO",
    reportNumber = "MIT-CTP/5963",
    month = "11",
    year = "2025"
}

@article{Toomey:2025xyo,
    author = "Toomey, Michael W. and Montefalcone, Gabriele and McDonough, Evan and Freese, Katherine",
    title = "{How theory-informed priors affect DESI evidence for evolving dark energy}",
    eprint = "2509.13318",
    archivePrefix = "arXiv",
    primaryClass = "astro-ph.CO",
    reportNumber = "MIT-CTP/5923, UTWI-29-2025, NORDITA-2025-049",
    doi = "10.1103/snyr-qs56",
    journal = "Phys. Rev. D",
    volume = "113",
    number = "12",
    pages = "123532",
    year = "2026"
}

@article{Alexander:2022own,
    author = "Alexander, Stephon and Bernardo, Heliudson and Toomey, Michael W.",
    title = "{Addressing the Hubble and S $_{8}$ tensions with a kinetically mixed dark sector}",
    eprint = "2207.13086",
    archivePrefix = "arXiv",
    primaryClass = "astro-ph.CO",
    doi = "10.1088/1475-7516/2023/03/037",
    journal = "JCAP",
    volume = "03",
    pages = "037",
    year = "2023"
}

@article{Brookfield:2005bz,
    author = "Brookfield, Anthony W. and van de Bruck, C. and Mota, D. F. and Tocchini-Valentini, D.",
    title = "{Cosmology of mass-varying neutrinos driven by quintessence: theory and observations}",
    eprint = "astro-ph/0512367",
    archivePrefix = "arXiv",
    doi = "10.1103/PhysRevD.73.083515",
    journal = "Phys. Rev. D",
    volume = "73",
    pages = "083515",
    year = "2006",
    note = "[Erratum: Phys.Rev.D 76, 049901 (2007)]"
}

@article{Garcia-Bellido:1992xlz,
    author = "Garcia-Bellido, Juan",
    title = "{Dark matter with variable masses}",
    eprint = "hep-ph/9205216",
    archivePrefix = "arXiv",
    reportNumber = "IEM-FT-54-92",
    doi = "10.1142/S0218271893000076",
    journal = "Int. J. Mod. Phys. D",
    volume = "2",
    pages = "85--95",
    year = "1993"
}

@inproceedings{Anderson:1997un,
    author = "Anderson, Greg W. and Carroll, Sean M.",
    title = "{Dark matter with time dependent mass}",
    booktitle = "{1st International Conference on Particle Physics and the Early Universe}",
    eprint = "astro-ph/9711288",
    archivePrefix = "arXiv",
    reportNumber = "NSF-ITP-97-146, NUHEP-TH-97-14",
    doi = "10.1142/9789814447263_0025",
    pages = "227--229",
    month = "9",
    year = "1997"
}

@article{Das:2023enn,
    author = "Das, Anirban and Das, Subinoy and Sethi, Shiv K.",
    title = "{Mass-varying dark matter and its cosmological signature}",
    eprint = "2303.17947",
    archivePrefix = "arXiv",
    primaryClass = "astro-ph.CO",
    reportNumber = "SLAC-PUB-17709",
    doi = "10.1103/PhysRevD.108.083501",
    journal = "Phys. Rev. D",
    volume = "108",
    number = "8",
    pages = "083501",
    year = "2023"
}

@article{Amendola:2000uh,
    author = "Amendola, Luca and Tocchini-Valentini, Domenico",
    title = "{Stationary dark energy: The Present universe as a global attractor}",
    eprint = "astro-ph/0011243",
    archivePrefix = "arXiv",
    doi = "10.1103/PhysRevD.64.043509",
    journal = "Phys. Rev. D",
    volume = "64",
    pages = "043509",
    year = "2001"
}

@article{Farrar:2003uw,
    author = "Farrar, Glennys R. and Peebles, P. James E.",
    title = "{Interacting dark matter and dark energy}",
    eprint = "astro-ph/0307316",
    archivePrefix = "arXiv",
    doi = "10.1086/381728",
    journal = "Astrophys. J.",
    volume = "604",
    pages = "1--11",
    year = "2004"
}

@article{Casas:1991ky,
    author = "Casas, J. A. and Garcia-Bellido, J. and Quiros, M.",
    title = "{Scalar - tensor theories of gravity with phi dependent masses}",
    eprint = "hep-ph/9204213",
    archivePrefix = "arXiv",
    reportNumber = "IEM-FT-40-91",
    doi = "10.1088/0264-9381/9/5/018",
    journal = "Class. Quant. Grav.",
    volume = "9",
    pages = "1371--1384",
    year = "1992"
}

@article{Fardon:2003eh,
    author = "Fardon, Rob and Nelson, Ann E. and Weiner, Neal",
    title = "{Dark energy from mass varying neutrinos}",
    eprint = "astro-ph/0309800",
    archivePrefix = "arXiv",
    reportNumber = "UW-PT-03-22",
    doi = "10.1088/1475-7516/2004/10/005",
    journal = "JCAP",
    volume = "10",
    pages = "005",
    year = "2004"
}

@article{Mandal:2022yym,
    author = "Mandal, Sayan and Sehgal, Neelima",
    title = "{Mass-varying dark matter from a phase transition}",
    eprint = "2212.07884",
    archivePrefix = "arXiv",
    primaryClass = "hep-ph",
    doi = "10.1103/PhysRevD.107.123003",
    journal = "Phys. Rev. D",
    volume = "107",
    number = "12",
    pages = "123003",
    year = "2023"
}

@article{Wang:2016lxa,
    author = "Wang, B. and Abdalla, E. and Atrio-Barandela, F. and Pavon, D.",
    title = "{Dark Matter and Dark Energy Interactions: Theoretical Challenges, Cosmological Implications and Observational Signatures}",
    eprint = "1603.08299",
    archivePrefix = "arXiv",
    primaryClass = "astro-ph.CO",
    doi = "10.1088/0034-4885/79/9/096901",
    journal = "Rept. Prog. Phys.",
    volume = "79",
    number = "9",
    pages = "096901",
    year = "2016"
}

@article{SevillanoMunoz:2024ayh,
    author = "Sevillano Mu{\~n}oz, Sergio",
    title = "{A particle's perspective on screening mechanisms}",
    eprint = "2407.08779",
    archivePrefix = "arXiv",
    primaryClass = "hep-ph",
    reportNumber = "IPPP/24/45",
    doi = "10.1088/1475-7516/2024/12/052",
    journal = "JCAP",
    volume = "12",
    pages = "052",
    year = "2024"
}

@article{Trodden:2022zye,
    author = "Trodden, Mark",
    title = "{Coupled Early Dark Energy}",
    eprint = "2209.15046",
    archivePrefix = "arXiv",
    primaryClass = "astro-ph.CO",
    doi = "10.1007/978-3-031-38477-6_12",
    journal = "Springer Proc. Phys.",
    volume = "392",
    pages = "223--236",
    year = "2024"
}

@article{Freedman:2020dne,
    author = "Freedman, Wendy L. and Madore, Barry F. and Hoyt, Taylor and Jang, In Sung and Beaton, Rachael and Lee, Myung Gyoon and Monson, Andrew and Neeley, Jill and Rich, Jeffrey",
    title = "{Calibration of the Tip of the Red Giant Branch}",
    eprint = "2002.01550",
    archivePrefix = "arXiv",
    primaryClass = "astro-ph.GA",
    doi = "10.3847/1538-4357/ab7339",
    journal = "Astrophys. J.",
    volume = "891",
    number = "1",
    pages = "57",
    year = "2020"
}

@article{H0LiCOW:2019pvv,
    author = "Wong, Kenneth C. and others",
    collaboration = "H0LiCOW",
    title = "{H0LiCOW {\textendash} XIII. A 2.4 per cent measurement of H0 from lensed quasars: 5.3{\ensuremath{\sigma}} tension between early- and late-Universe probes}",
    eprint = "1907.04869",
    archivePrefix = "arXiv",
    primaryClass = "astro-ph.CO",
    doi = "10.1093/mnras/stz3094",
    journal = "Mon. Not. Roy. Astron. Soc.",
    volume = "498",
    number = "1",
    pages = "1420--1439",
    year = "2020"
}

@article{Birrer:2020tax,
    author = "Birrer, S. and others",
    title = "{TDCOSMO - IV. Hierarchical time-delay cosmography {\textendash} joint inference of the Hubble constant and galaxy density profiles}",
    eprint = "2007.02941",
    archivePrefix = "arXiv",
    primaryClass = "astro-ph.CO",
    doi = "10.1051/0004-6361/202038861",
    journal = "Astron. Astrophys.",
    volume = "643",
    pages = "A165",
    year = "2020"
}

@article{Nesseris:2025lke,
    author = "Nesseris, Savvas and Akrami, Yashar and Starkman, Glenn D.",
    title = "{To CPL, or not to CPL? What we have not learned about the dark energy equation of state}",
    eprint = "2503.22529",
    archivePrefix = "arXiv",
    primaryClass = "astro-ph.CO",
    reportNumber = "IFT-UAM/CSIC-25-29",
    month = "3",
    year = "2025"
}

@article{Pan:2025qwy,
    author = "Pan, Supriya and Paul, Sivasish and Saridakis, Emmanuel N. and Yang, Weiqiang",
    title = "{Interacting dark energy after DESI DR2: A challenge for the {\ensuremath{\Lambda}}CDM paradigm?}",
    eprint = "2504.00994",
    archivePrefix = "arXiv",
    primaryClass = "astro-ph.CO",
    doi = "10.1103/5y21-k39n",
    journal = "Phys. Rev. D",
    volume = "113",
    number = "2",
    pages = "023515",
    year = "2026"
}

@article{deSouza:2025rhv,
    author = "de Souza, Rayff and Rodrigues, Gabriel and Alcaniz, Jailson",
    title = "{Thawing quintessence and transient cosmic acceleration in light of DESI}",
    eprint = "2504.16337",
    archivePrefix = "arXiv",
    primaryClass = "astro-ph.CO",
    doi = "10.1103/2tjq-dtbc",
    journal = "Phys. Rev. D",
    volume = "112",
    number = "8",
    pages = "083533",
    year = "2025"
}

@article{Carroll:2003st,
    author = "Carroll, Sean M. and Hoffman, Mark and Trodden, Mark",
    title = "{Can the dark energy equation-of-state parameter $w$ be less than $-1$?}",
    eprint = "astro-ph/0301273",
    archivePrefix = "arXiv",
    reportNumber = "EFI-2003-01, SU-GP-03-1-1",
    doi = "10.1103/PhysRevD.68.023509",
    journal = "Phys. Rev. D",
    volume = "68",
    pages = "023509",
    year = "2003"
}

@article{Cline:2003gs,
    author = "Cline, James M. and Jeon, Sangyong and Moore, Guy D.",
    title = "{The Phantom menaced: Constraints on low-energy effective ghosts}",
    eprint = "hep-ph/0311312",
    archivePrefix = "arXiv",
    reportNumber = "MCGILL-03-25",
    doi = "10.1103/PhysRevD.70.043543",
    journal = "Phys. Rev. D",
    volume = "70",
    pages = "043543",
    year = "2004"
}

@article{Dubovsky:2005xd,
    author = "Dubovsky, S. and Gregoire, T. and Nicolis, A. and Rattazzi, R.",
    title = "{Null energy condition and superluminal propagation}",
    eprint = "hep-th/0512260",
    archivePrefix = "arXiv",
    reportNumber = "CERN-PH-TH-2005-265, HUTP-05-A0056",
    doi = "10.1088/1126-6708/2006/03/025",
    journal = "JHEP",
    volume = "03",
    pages = "025",
    year = "2006"
}

@article{Creminelli:2012my,
    author = "Creminelli, Paolo and Hinterbichler, Kurt and Khoury, Justin and Nicolis, Alberto and Trincherini, Enrico",
    title = "{Subluminal Galilean Genesis}",
    eprint = "1209.3768",
    archivePrefix = "arXiv",
    primaryClass = "hep-th",
    doi = "10.1007/JHEP02(2013)006",
    journal = "JHEP",
    volume = "02",
    pages = "006",
    year = "2013"
}

@article{Nicolis:2009qm,
    author = "Nicolis, Alberto and Rattazzi, Riccardo and Trincherini, Enrico",
    title = "{Energy's and amplitudes' positivity}",
    eprint = "0912.4258",
    archivePrefix = "arXiv",
    primaryClass = "hep-th",
    doi = "10.1007/JHEP05(2010)095",
    journal = "JHEP",
    volume = "05",
    pages = "095",
    year = "2010",
    note = "[Erratum: JHEP 11, 128 (2011)]"
}

@article{Creminelli:2010ba,
    author = "Creminelli, Paolo and Nicolis, Alberto and Trincherini, Enrico",
    title = "{Galilean Genesis: An Alternative to inflation}",
    eprint = "1007.0027",
    archivePrefix = "arXiv",
    primaryClass = "hep-th",
    doi = "10.1088/1475-7516/2010/11/021",
    journal = "JCAP",
    volume = "11",
    pages = "021",
    year = "2010"
}

@article{Linder:2025zxb,
    author = "Linder, Eric V.",
    title = "{Uplifting, Depressing, and Tilting Dark Energy}",
    eprint = "2506.02122",
    archivePrefix = "arXiv",
    primaryClass = "astro-ph.CO",
    month = "6",
    year = "2025"
}

@article{Agrawal:2019dlm,
    author = "Agrawal, Prateek and Obied, Georges and Vafa, Cumrun",
    title = "{$H_0$ tension, swampland conjectures, and the epoch of fading dark matter}",
    eprint = "1906.08261",
    archivePrefix = "arXiv",
    primaryClass = "astro-ph.CO",
    doi = "10.1103/PhysRevD.103.043523",
    journal = "Phys. Rev. D",
    volume = "103",
    number = "4",
    pages = "043523",
    year = "2021"
}

@article{Bedroya:2025fwh,
    author = "Bedroya, Alek and Obied, Georges and Vafa, Cumrun and Wu, David H.",
    title = "{Evolving Dark Sector and the Dark Dimension Scenario}",
    eprint = "2507.03090",
    archivePrefix = "arXiv",
    primaryClass = "astro-ph.CO",
    month = "7",
    year = "2025"
}

@article{Lee:2026yzs,
    author = "Lee, Dong Ha and van de Bruck, Carsten and Di Valentino, Eleonora and Van Waerbeke, Ludovic and Zhitnitsky, Ariel",
    title = "{Evolving Dark Energy Is Vacuum Energy After All}",
    eprint = "2606.20036",
    archivePrefix = "arXiv",
    primaryClass = "astro-ph.CO",
    month = "6",
    year = "2026"
}

@article{Khoury:2026svx,
    author = "Khoury, Justin and Lin, Meng-Xiang and Trodden, Mark",
    title = "{Cosmological Evidence for Dark Axion-Dark Baryon Interactions from Apparent Phantom Crossing}",
    eprint = "2607.16191",
    archivePrefix = "arXiv",
    primaryClass = "astro-ph.CO",
    month = "7",
    year = "2026"
}

@article{Pitrou:2023swx,
    author = "Pitrou, Cyril and Uzan, Jean-Philippe",
    title = "{Hubble Tension as a Window on the Gravitation of the Dark Matter Sector}",
    eprint = "2312.12493",
    archivePrefix = "arXiv",
    primaryClass = "astro-ph.CO",
    doi = "10.1103/PhysRevLett.132.191001",
    journal = "Phys. Rev. Lett.",
    volume = "132",
    number = "19",
    pages = "191001",
    year = "2024"
}

@article{Archidiacono:2022iuu,
    author = "Archidiacono, Maria and Castorina, Emanuele and Redigolo, Diego and Salvioni, Ennio",
    title = "{Unveiling dark fifth forces with linear cosmology}",
    eprint = "2204.08484",
    archivePrefix = "arXiv",
    primaryClass = "astro-ph.CO",
    reportNumber = "CERN-TH-2022-066",
    doi = "10.1088/1475-7516/2022/10/074",
    journal = "JCAP",
    volume = "10",
    pages = "074",
    year = "2022"
}

@article{Hees:2018fpg,
    author = "Hees, Aur{\'e}lien and Minazzoli, Olivier and Savalle, Etienne and Stadnik, Yevgeny V. and Wolf, Peter",
    title = "{Violation of the equivalence principle from light scalar dark matter}",
    eprint = "1807.04512",
    archivePrefix = "arXiv",
    primaryClass = "gr-qc",
    doi = "10.1103/PhysRevD.98.064051",
    journal = "Phys. Rev. D",
    volume = "98",
    number = "6",
    pages = "064051",
    year = "2018"
}

@article{Lee:2020zjt,
    author = "Lee, J. G. and Adelberger, E. G. and Cook, T. S. and Fleischer, S. M. and Heckel, B. R.",
    title = "{New Test of the Gravitational $1/r^2$ Law at Separations down to 52 $\mu$m}",
    eprint = "2002.11761",
    archivePrefix = "arXiv",
    primaryClass = "hep-ex",
    doi = "10.1103/PhysRevLett.124.101101",
    journal = "Phys. Rev. Lett.",
    volume = "124",
    number = "10",
    pages = "101101",
    year = "2020"
}

@article{Bean:2008ac,
    author = "Bean, Rachel and Flanagan, Eanna E. and Laszlo, Istvan and Trodden, Mark",
    title = "{Constraining Interactions in Cosmology's Dark Sector}",
    eprint = "0808.1105",
    archivePrefix = "arXiv",
    primaryClass = "astro-ph",
    doi = "10.1103/PhysRevD.78.123514",
    journal = "Phys. Rev. D",
    volume = "78",
    pages = "123514",
    year = "2008"
}

@article{Li:2026xaz,
    author = "Li, Tian-Nuo and Giar{\`e}, William and Du, Guo-Hong and Li, Yun-He and Di Valentino, Eleonora and Zhang, Jing-Fei and Zhang, Xin",
    title = "{Robust Preference for Dark Sector Interactions}",
    eprint = "2601.07361",
    archivePrefix = "arXiv",
    primaryClass = "astro-ph.CO",
    month = "1",
    year = "2026"
}

@article{Baidya:2026tyh,
    author = "Baidya, Mainak and Christiansen, {\O}yvind and da Fonseca, Vitor and Linder, Eric V. and Mota, David F.",
    title = "{Mass-Varying Neutrinos from an Inverse Symmetron}",
    eprint = "2606.07391",
    archivePrefix = "arXiv",
    primaryClass = "astro-ph.CO",
    month = "6",
    year = "2026"
}

@article{vandeBruck:2019vzd,
    author = "van de Bruck, Carsten and Thomas, Cameron C.",
    title = "{Dark Energy, the Swampland and the Equivalence Principle}",
    eprint = "1904.07082",
    archivePrefix = "arXiv",
    primaryClass = "hep-th",
    doi = "10.1103/PhysRevD.100.023515",
    journal = "Phys. Rev. D",
    volume = "100",
    number = "2",
    pages = "023515",
    year = "2019"
}

@article{Brax:2011ja,
    author = "Brax, Philippe and van de Bruck, Carsten and Davis, Anne-Christine and Li, Baojiu and Shaw, Douglas J.",
    title = "{Nonlinear Structure Formation with the Environmentally Dependent Dilaton}",
    eprint = "1102.3692",
    archivePrefix = "arXiv",
    primaryClass = "astro-ph.CO",
    doi = "10.1103/PhysRevD.83.104026",
    journal = "Phys. Rev. D",
    volume = "83",
    pages = "104026",
    year = "2011"
}

@article{Kesden:2006vz,
    author = "Kesden, Michael and Kamionkowski, Marc",
    title = "{Tidal Tails Test the Equivalence Principle in the Dark Sector}",
    eprint = "astro-ph/0608095",
    archivePrefix = "arXiv",
    doi = "10.1103/PhysRevD.74.083007",
    journal = "Phys. Rev. D",
    volume = "74",
    pages = "083007",
    year = "2006"
}

\end{document}